\documentclass[11pt]{article}

\usepackage{amsmath}

\newcommand{\remove}[1]{}

\newtheorem{theorem}{Theorem}

\newtheorem{definition}{Definition}

\newtheorem{claim}{Claim}
\newtheorem{Lemma}[claim]{Lemma}

\newtheorem{corollary}{Corollary}
\newtheorem{fact}{Fact}

\newcommand{\qedsymb}{\hfill{\rule{2mm}{2mm}}}
\newenvironment{proof}[1][]{\begin{trivlist}
\item[\hspace{\labelsep}{\bf\noindent Proof#1:\/}] }{\qedsymb\end{trivlist}}

\def\H{{\cal H}}
\def\ra{\rangle}
\def\rket{\ra}
\def\la{\langle}
\def\lbra{\la}
\def\up{\Uparrow}
\def\dn{\Downarrow}
\def\lt{\Leftarrow}
\def\rt{\Rightarrow}

\newcommand{\be}{\begin{eqnarray}}
\newcommand{\ee}{\end{eqnarray}}

\renewcommand{\epsilon}{\varepsilon}

\newcommand{\ket}[1]{|#1\rangle}
\newcommand{\bra}[1]{\langle#1|}

\newsavebox{\fmbox}
\newenvironment{fmpage}[1]
     {\medskip\begin{lrbox}{\fmbox}\begin{minipage}{#1}}
     {\end{minipage}\end{lrbox}\fbox{\usebox{\fmbox}}\medskip}

\begin{document}

\title{Coins Make Quantum Walks Faster}

\author{
Andris Ambainis\\
School of Mathematics, \\
Institute for Advanced Study, \\
Princeton, NJ 08540\\
ambainis@ias.edu
\and
 Julia Kempe\\
CNRS-LRI UMR 8623\\
 Universit\'e de Paris-Sud\\ 91405 Orsay, France \\
and UC Berkeley, Berkeley, CA 94720\\
kempe@lri.fr
 \and
Alexander Rivosh\\
Institute of Mathematics and Computer Science\\
University of Latvia\\
Raina bulv.29, Riga, LV-1459, Latvia\\
vbug@solutions.lv
}

\date{\today}
\maketitle

\begin{abstract}
We show how to search $N$ items arranged on a $\sqrt{N}\times\sqrt{N}$ 
grid in time $O(\sqrt N \log N)$, using a discrete time quantum walk.
This result for the first time exhibits a significant difference
between discrete time and continuous time walks without coin degrees
of freedom, since it has been shown recently that such a continuous
time walk needs time $\Omega(N)$ to perform the same task. Our result
furthermore improves on a previous bound for quantum local search by
Aaronson and Ambainis. We generalize our result to $3$ and more
dimensions where the walk yields the optimal performance of
$O(\sqrt{N})$ and give several extensions of quantum walk search
algorithms for general graphs. The coin-flip operation needs to be
chosen judiciously: we show that another ``natural'' choice of coin
gives a walk that takes $\Omega(N)$ steps. We also show that in $2$
dimensions it is sufficient to have a two-dimensional coin-space to
achieve the time $O(\sqrt{N} \log N)$.
\end{abstract}

\section{Introduction}

Quantum walks are quantum counterparts of classical random walks.
Classical random walks have many applications in randomized algorithms
\cite{Motwani:book} and we hope that quantum walks would have
similar applications in quantum algorithms.  Both
discrete-time \cite{Meyer:96a,Aharonov:01a,Ambainis:01b} and continous
time \cite{Farhi:98a,Childs:01a} quantum walks have been
introduced\footnote{For an introduction to quantum walks see
\cite{Kempe:03b}.}. The definitions of the two are quite different.
In continous time, one can directly define the walk on the vertices of
the graph. In discrete time, it is necessary to introduce an extra
``coin'' register storing the direction in which the walk is
moving. 

Because of this difference in the definitions, it has been open what
the relation between discrete and continous walk is. In the classical
world, the continous walk is the limit of the discrete walk when the
length of the time step approaches 0. In the quantum case, this is no
longer true. Even if we make the time-steps of the discrete walk
smaller and smaller, the ``coin'' register remains.  Therefore, the
limit cannot be the continous walk without the ``coin'' register. This
means that one variant of quantum walks could be more powerful than
the other in some context, but so far all known examples have given
similar behavior of the two walks (see
e.g. \cite{Childs:01a,Kempe:02b,Childs:02a}).

In this paper, we present the first example where the discrete walk
(with ``coin'') outperforms the continous walk (with no ``coin'').
Our example is the spatial search \cite{Benioff:02a,Aaronson:03a}
variant of Grover's search problem.  In the usual Grover's search
problem \cite{Grover:96a}, we have $N$ items, one of which is
marked. Then, we can find the marked item in $O(\sqrt{N})$ quantum
steps, with one quantum step querying a superposition of items. In
contrast, classically $\Omega(N)$ queries are required.
%
In the ``spatial search" variant, we have the extra constraint
that the $N$ items are stored in $N$ different memory
locations and we need time to move betwen locations. 
This may increase the running time of a quantum algorithm.

The first ``spatial'' version of Grover's algorithms with optimal
performance was given by \cite{Shenvi:02b} who showed how to search
$N$ items arranged on the $n$-dimensional hypercube, using a
discrete quantum walk.   

In this paper, we consider the $2$-dimensional arrangement 
where $N$ memory locations are arranged in an $\sqrt{N}\times \sqrt{N}$
grid. 
This was first studied by Benioff \cite{Benioff:02a} who observed that the
usual Grover's search algorithm takes $\Omega(N)$ steps. It uses
$\Theta(\sqrt{N})$ query steps but, between each two queries, it might
move a distance of $\Theta(\sqrt{N})$. Thus, the total time becomes
$\Theta(N)$ and the quantum speedup disappears.  Aaronson and Ambainis
\cite{Aaronson:03a} fixed this problem by giving an algorithm for
searching the $2$-dimensional grid in $O(\sqrt{N}\log^2 N)$ total
steps\footnote{The running times for the $2$-dimensional grid are
for the case when the grid contains one marked item.
The general case (an arbitrary number of marked items) can be reduced
to the one item case with a $\log N$ increase of the running time \cite{Aaronson:03a}.
That would result in a running time of $O(\sqrt{N} \log^3 N)$ for the
algorithm of \cite{Aaronson:03a} and $O(\sqrt{N} \log^2 N)$ for
our algorithm which we present in this paper.
For 3 and higher dimensions, the general case can be reduced to
the one item case with just a constant factor increase \cite{Aaronson:03a}.
Thus, the asymptotic running times stay the same.}
 (counting both queries and moving steps) and the $3$-dimensional
grid in $O(\sqrt{N})$ steps, using Grover's algorithm together with
multilevel recursion.  Quantum walks were first applied to this
problem by Childs and Goldstone \cite{Childs:03a} who
studied the search on the grid by a continous quantum walk. They
discovered that the continous walk provides an alternative search
algorithm with optimal performance of $O(\sqrt{N})$ in $5$ and more
dimensions, but not in $2$ or $3$ dimensions, where the continuous
walk takes $\Omega(N)$ and $\Omega(N^{5/6})$, respectively. 
In $4$ dimensions the continuous time walk algorithm
performs as $O(\sqrt{N} \log N)$. 

In this paper, we use discrete-time quantum walks to design an
algorithm that searches the grid in $O(\sqrt{N}\log N)$ time in $2$
dimensions and $O(\sqrt{N})$ time in $3$ and more dimensions.  Thus, our
algorithm is faster than both the non-walk quantum algorithm of
\cite{Aaronson:03a} and the algorithm based on the continous time quantum
walk \cite{Childs:03a}. 
 In addition to having a very simple structure our algorithm also
uses only 1 or 2 qubits of extra memory (or $log 2d$ qubits for the d-dimensional grid), besides the current location.
(The previous algorithm of \cite{Aaronson:03a} uses $O(\log^c n)$ qubits of extra memory.)

Besides improving the running time, 
we present
several interesting features of quantum walks.  The first feature is
that the discrete-time walk succeeds while the continous walk does not.
Secondly, the behavior of the discrete quantum walk on the grid crucially
depends on the choice of the ``coin'' transformation. 
One natural choice,  discovered numerically by Neil Shenvi \cite{Shenvi:unp},
leads to our algorithm while some other natural choices fail to
produce a good algorithm. Thus, the ``coin'' transformation could be a
resource which affects the algorithm profoundly. We give both upper and lower bounds for the performance of some natural choices of the ``coin''. Surprisingly we show that in the case of the $2$-dimensional grid only $2$ (and not the standard $4$) coin-degrees of freedom are sufficient to achieve the quantum speed-up. The insights gained
from our study might aid in the design of future discrete quantum walk
based algorithms. Several such algorithms have recently been
discovered \cite{Childs:02a,Ambainis:03a,Magniez:03a,Childs:03b,Szegedy:04a}.

Our presentation allows a fairly general approach to quantum walk
search algorithms on graphs. In particular it simplifies the proof of
\cite{Shenvi:02b}, where the relevant eigenvectors had to be
``guessed''. We also give a discrete walk search algorithm on the
complete graph and show its equivalence to Grover's algorithm and
outline several generalizations of our results.

\section{Preliminaries and Notation}

\subsection{Model}

Our model is similar to the one in \cite{Aaronson:03a}.
We have an {\em undirected graph} $G=(V,E)$. 
Each vertex $v$ stores a variable $a_v\in\{0, 1\}$.
Our goal is to find a vertex $v$ for which $a_v=1$ (assuming
such vertex exists). We will often call such vertices marked
and vertices for which $a_v=0$ unmarked.

In one step, an algorithm can examine the current vertex or move to
a neighboring vertex in the graph $G$.
The goal is to find a marked vertex in as few steps as possible.


More formally, a quantum algorithm is a sequence of 
unitary transformations on a Hilbert space
$\H_i\otimes \H_V$. $\H_V$ is a Hilbert space spanned by
states $\ket{v}$ corresponding to vertices of $G$.
$\H_i$ represents the algorithm's internal state and can be of
arbitrary fixed dimension.
A $t$-step quantum algorithm is a sequence 
$U_1$, $U_2$, $\ldots$, $U_t$ where each $U_i$ is either a {\em query}
or a {\em local transformation}. A query $U_i$ consists of two transformations
($U_i^0$, $U_i^1$). $ U_i^0  \otimes I$ is applied to 
all $\H_i \otimes |v\ra $ for which $a_{v}=0$
and $ U_i^1 \otimes I$ is applied to 
all $\H_i \otimes |v\ra $ for which $a_{v}=1$.

A local transformation can be defined in several ways \cite{Aaronson:03a}.
In this paper, we require them to be $Z$-local.
A transformation $U_i$ is $Z$-local if, for any $v\in V$ and
$\ket{\psi}\in \H_i$, the state $U_i(\ket{\psi}\otimes \ket{v})$
is contained in the subspace $\H_i\otimes \H_{\Gamma(v)}$ where
$\H_{\Gamma(v)} \subset \H_V$ is spanned by the state $\ket{v}$ and the states
$\ket{v'}$ for all $v'$ adjacent to $v$.
Our results also apply if the local transformations are $C$-local
(another locality definition introduced in \cite{Aaronson:03a}).

The algorithm starts in a fixed starting state $\ket{\psi_{start}}$ and
applies $U_1$, $\ldots$, $U_t$. This results in a final state
$\ket{\psi_{final}}=U_t U_{t-1} \ldots U_1 \ket{\psi_{start}}$.
Then, we measure $\ket{\psi_{start}}$.
The algorithm succeeds if measuring the $\H_V$ part of the 
final state gives $\ket{g}$ such that $a_{g}=1$. 
 
For more details on this model, see \cite{Aaronson:03a}.

\subsection{Search by quantum walk}

In what follows we will assume that $G$ is
{\em undirected} and {\em $d$-regular}, i.e. has constant degree
$d$. To each vertex we can associate a labeling $\{1,\ldots,d\}$ of
the $d$ edges (directions) adjacent to it and an auxiliary
``coin''-Hilbert space ${\cal{H}}_d=\{|1\ra,\ldots,|d\ra\}$. Let
${\cal{H}}_N$ be the Hilbert space spanned by the vertices of the
graph, then the walk takes place in the joint space of coin and graph
${\cal{H}}={\cal{H}}_d \otimes {\cal H}_N$.

\begin{definition}{[Discrete Quantum Walk on G:]} The discrete quantum walk is an alternation of coin flip and moving step: $U=S \cdot C$, where  $S$ is a shift  controlled by the coin register 
\begin{equation} \label{eq:shift}
S: \, |i\ra \otimes |x\ra \longrightarrow |\pi(i)\ra \otimes |\tilde{x}\ra
\end{equation}
$ i=1, \ldots , d$ and $ x,\tilde{x} \in V$, $x$ and $\tilde{x}$ are connected by the edge labelled ``$i$'' on $x$'s side and $\pi$ is a permutation of the $d$ basis states of the coin space  ${\cal H}_d$, and the coin $C=C_0\otimes I_N$
where $I_N$ acts as identity on ${\cal{H}}_N$ and $C_0$ is a ``coin-flip'' acting on ${\cal{H}}_d$ 
\begin{equation}
C_0=2|s\ra \la s|-I_d \quad \mathtt{where} \quad |s\ra =\frac{1}{\sqrt{d}} \sum_{i=1}^d |i\ra. 
\end{equation}
\end{definition}
For a given $i$ $S$ permutes the vertices of the graph, hence $S$ is a unitary operation. The permutation $\pi$ allows us to specify shift operations that act differently on the coin space  ${\cal H}_d$.
Note that the coin is {\em symmetric} in that it treats all $d$ directions equally, and among all such coins $C_0$ is the one farthest away from identity.

{\em Remark:} The uniform superposition $|\Phi_0 \ra=\frac{1}{\sqrt{dN}}\sum_{i=1}^d \sum_{x=1}^N |i\ra \otimes |x\ra$ is an eigenvector of $U$ with eigenvalue $1$ ($U |\Phi_0
\ra=|\Phi_0 \ra$); if we start the walk in $|\Phi_0\ra$ it will never change this state.

To introduce a marked item in the graph we need to have an inhomogeneity in the quantum walk by using the coin to ``mark'' a vertex $v$, which gives rise to the following:

\begin{definition}{[Perturbed Quantum Walk:]} The perturbed walk with marked vertex $v$ and ``marking coin'' $C_1=-I_d$ is $U'=S \cdot C'$, where 
\begin{equation}
C'=C_0 \otimes (I-|v\ra \la v|)+C_1 \otimes |v\ra \la v|=C-(C_0-C_1) \otimes |v\ra \la v|.
\end{equation}
\end{definition}
We will think of $U'$ as the random walk with one (or several)
marked coins. This means that instead of one coin for all nodes, $C_0
\otimes I$, we have a different coin $C_1$ on the marked state. Numerical data shows that other marked
coins exhibit similar properties as $C_1=-I$, but we will use this coin
which simplifies the analysis.  Then $C_0-C_1=2 |s\ra \la s|$, and $U'=U-2 S   |s, v\ra \la s, v|=U\cdot (I_{dN}- 2 |s, v\ra \la s, v|)$ using $C_0 |s\ra=|s\ra$.

The quantum walk $U$ gives rise to a search algorithm on a graph $G$ in the following way:

\begin{center}
\begin{fmpage}{8cm}
\textbf{Quantum Walk Search Algorithm}
\begin{enumerate}
\item Initialise the quantum system in the uniform superposition $|\Phi_0\ra$.
\item Do $T$ times:
 Apply the marked walk $U'$.
\item Measure the position register.
\item Check if the measured vertex 
is the marked item.
\end{enumerate}
\end{fmpage}
\end{center}
An item on a vertex of the graph could be marked by setting an auxiliary qubit to $|1\ra$, whereas the unmarked items could have this qubit set to $|0\ra$. Then this auxiliary qubit can control the coin to be $C$ for the unmarked items and $C'$ for the marked item. 

We will analyse this algorithm to obtain upper bounds on the query
complexity of search by random walks.
\paragraph{Complete Graph - Grover's Algorithm:} As a first example let us illustrate how we can view  Grover's
algorithm \cite{Grover:96a} as a random walk search algorithm on the complete
graph. Each vertex has $N$ edges (we will include a self-loop for each vertex). Both vertices and edges are labelled with $1, \ldots , N$; the coin space and the vertex Hilbert space are both $N$-dimensional and we will write states as $|i\ra \otimes |j\ra$, where the first register is the coin-register. The
shift operation $S$ is defined as
$$S:\, |i\ra \otimes |j\ra \longrightarrow |j\ra \otimes |i\ra.$$
The marked coin in this case is chosen to be
$C_1=-C_0$, which gives $C_1-C_0=-2C_0$ and $C'=C_0 \otimes (I-2
|v\ra \la v|)$, where $|v\ra$ is the marked
state. Note that $C_0=2|s\ra \la s|-1_N$ is the reflection around the mean operator of Grover's (``standard'') algorithms and
$I-2 |v\ra \la v|)=:R_v$ the phase flip of the
oracle. Recall that Grover's algorithm is of the form $(R_v \cdot C_0)^T|s\ra$. The initial state for the random walk based algorithm is the uniform superposition $|\Phi_0\ra = |s\ra \otimes |s\ra$. Now $U'|\Phi_0\ra =S \cdot C' |\Phi_0\ra=R_v|s\ra \otimes C_0|s\ra$, $C'\cdot U'|\Phi_0\ra=(C_0 \cdot
R_v)|s\ra \otimes (R_v \cdot C_0)|s\ra$ and $U'^2|\Phi_0\ra= (R_v \cdot C_0)|s\ra \otimes (C_0 \cdot R_v)|s\ra$. So we see that a random walk in this scenario gives
exactly Grovers algorithm on both the coin space and the vertex space,
at the expense of a factor of 2 in the number of applications.

\section{Results in $2$ dimensions}

We give several upper and lower bounds for the discrete quantum walk
on the grid. The $N$ memory locations are arranged in a $\sqrt{N} \times \sqrt{N}$ grid $G$, 
labeled by their $x$ and $y$ coordinate as $|x,y\ra$ for 
$x,y \in \{0, \ldots ,\sqrt{N}-1\}$. 
will assume periodic boundary conditions and operate mod $\sqrt{N}$. The natural coin space is $4$-dimensional. We will label the edges emanating from each vertex with $\rightarrow, \leftarrow ,\uparrow, \downarrow$, indicating the positive and negative $x$ and $y$ directions. 

As it turns out, the choice of the coin transformation (or,
equivalently, of the permutation $\pi$ in Eq. (\ref{eq:shift})) is crucial
for the performance of the random walk. We will show that using a ``flip-flop'' shift, gives a search algorithm that succeeds in $O(\sqrt{N} \log N)$
time. The ``flip-flop'' shift $S_{ff}$ changes direction after every move, 
i.e. $\pi$ flips $\uparrow$ with $\downarrow$ and $\rightarrow$ with $\leftarrow$.
Our analysis of the ``flip-flop'' based walk follows the numerical discovery of its performance by Neil Shenvi \cite{Shenvi:unp}.
Another natural shift is the ``moving'' shift $S_m$ which does not change direction (i.e. in Eq. (\ref{eq:shift}) $\pi=id$ and $|\pi(i)\ra=|i\ra$).
\begin{equation} \label{eq:ff}
\begin{array}{rcrc}
S_{ff}: & |\rightarrow\ra \otimes |x,y\ra \longrightarrow |\leftarrow \ra \otimes |x+1,y\ra & \quad S_{m}: & |\rightarrow\ra \otimes |x,y\ra \longrightarrow |\rightarrow \ra \otimes |x+1,y\ra  \\
& |\leftarrow\ra \otimes |x,y\ra \longrightarrow |\rightarrow \ra \otimes |x-1,y\ra  && |\leftarrow\ra \otimes |x,y\ra \longrightarrow |\leftarrow \ra \otimes |x-1,y\ra  \\
& |\uparrow\ra \otimes |x,y\ra \longrightarrow |\downarrow \ra \otimes |x,y+1\ra && |\uparrow\ra \otimes |x,y\ra \longrightarrow |\uparrow \ra \otimes |x,y+1\ra  \\
& |\downarrow\ra \otimes |x,y\ra \longrightarrow |\uparrow \ra \otimes |x,y-1\ra && |\downarrow\ra \otimes |x,y\ra \longrightarrow |\downarrow \ra \otimes |x,y-1\ra 
\end{array} 
\end{equation}
Surprisingly we will show that the ``moving'' shift gives a walk search algorithm that takes time $\Omega(N)$. So even though it seems this walk ``moves faster'' than the ``flip-flop'' walk, the resulting algorithms performs much worse, no better than classical exhaustive search. It is this surprising behavior of $S_m$ which has halted the progress in finding a good discrete quantum walk search algorithm on the grid.

\begin{theorem}\label{Th:1}
For the quantum walk search algorithm associated to the quantum walk $U=S_{ff} \cdot C$, with $S_{ff}$ as in Eq. (\ref{eq:ff}),  there is a $T=O(\sqrt{N \log N})$, such that after $T$ steps the probability to determine the marked state is $p_T=O(1/\log N)$ .
\end{theorem}

\begin{corollary}\label{c:1}
We can get a local search algorithm based on the quantum walk that finds
the marked state with constant probability in time $O(\sqrt{N} \log N)$.
\end{corollary}

\begin{proof}[ of Corollary \ref{c:1}]
The initial state $|\Phi\ra$ can be generated with $\sqrt{N}$ local transformations.  Since we only have an estimate for $T$ up to a constant factor, we need to repeat the random walk an appropriate (constant) number of times. For the algorithm we will use amplitude amplification \cite{Brassard:02a} to achieve a time $O(\sqrt{N} \log N)$. We will give more details in the proof of Theorem \ref{Th:1}.

\end{proof}

\begin{theorem}\label{Th:2}
The quantum walk search algorithm associated with $S_m$ as in Eq. (\ref{eq:ff}) takes at least $\Omega(N)$ steps to  determine the marked state with constant probability.
\end{theorem}
We also consider a two dimensional coin inspired by Dirac's equation in 2+1 dimensions.
Let $\ket{\up}=\ket{0}$ and $\ket{\dn}=\ket{1}$ be the standard basis for
one qubit and $\ket{\lt}=\frac{1}{\sqrt{2}} \ket{0}+\frac{1}{\sqrt{2}} \ket{1}$
and $\ket{\rt}=\frac{1}{\sqrt{2}} \ket{0}-\frac{1}{\sqrt{2}} \ket{1}$
be the Hadamard basis. If there is no marked coins, one step of the quantum walk $U$ with the two-dimensional coin consists of:
\begin{enumerate}
\item
Move up/down:
\[\ket{\up} \otimes \ket{x} \otimes \ket{y} \rightarrow \ket{\up}\otimes \ket{x} \ket{y-1}, \]
\[\ket{\dn} \otimes \ket{x} \otimes \ket{y} \rightarrow \ket{\dn}\otimes \ket{x} \ket{y+1}. \]
\item
Move left/right:
\[\ket{\lt} \otimes \ket{x} \otimes \ket{y} \rightarrow \ket{\lt}\otimes \ket{x-1} \ket{y}, \]
\[\ket{\rt} \otimes \ket{x} \otimes \ket{y} \rightarrow \ket{\rt}\otimes \ket{x+1} \ket{y}. \]
\end{enumerate}
If there is a marked coin $\ket{v}$, we define the quantum walk
as $U'=U(I-2\ket{s,v}\bra{s,v})$ where $U$ is the walk with no marked coin and 
$\ket{s}$ is the state $\frac{1}{\sqrt{2}}\ket{\up}+\frac{1}{\sqrt{2}}\ket{\dn}$.

\begin{theorem}\label{Th:5}
The associated quantum walk search algorithm
takes $O(\sqrt{N \log N})$ steps  and the probability to measure the
marked state is $\Omega(1/\log N)$. This yields a local search algorithm running in time $O(\sqrt{N}\log N)$.
\end{theorem}

\section{Results in $3$ and more dimensions}

In more than $2$ dimensions the ``flip-flop'' based quantum walk search algorithms achieves its optimal performance of $O(\sqrt{N})$. Here $G$ is a grid of N vertices, arranged as $\sqrt[d]{N} \times \ldots \times \sqrt[d]{N}$, with periodic boundary conditions, as before, and states are labelled as $|x_1, \ldots , x_d\ra$. 

\begin{theorem}\label{Th:3}
Let $G$ be the $d$-dimensional grid with $N$ vertices.
Then the associated quantum walk with one marked coin takes
$O(\sqrt{N})$ steps and the probability to measure the marked state is
constant.
\end{theorem}

\begin{theorem}\label{Th:4}
The results of Theorems \ref{Th:1}, \ref{Th:2}, \ref{Th:5} and \ref{Th:3} hold also
for two marked items.
\end{theorem}

\section{Abstract search algorithm}
\label{sec:abstract}

Before giving the technical details let us give some intuition of the proof. Recall that Grover's algorithm in its standard form is a succession of reflections $R_v$ around the marked state $|v\ra$ followed by a reflection around the mean $R_{|\Phi\ra}=2|\Phi\ra \la \Phi|-I_N$, where $|\Phi\ra$ is the uniform superposition  over all items. It can be viewed as a rotation in a two dimensional space, spanned by the marked state $|v\ra$ and the initial state. In the basis where $|0\ra = |\Phi\ra$ and $|v\ra = \frac{1}{\sqrt{N}}|0\ra + \sqrt{\frac{N-1}{N}}|1\ra$, Grover's algorithms corresponds to the transformation (with $\sin \phi =2 \frac{\sqrt{N-1}}{N}$)
\begin{equation}\label{eq:rot}
U=\left(\begin{array}{cc} \cos \phi & -\sin \phi \\ \sin \phi & \cos \phi \end{array} \right).
\end{equation}
The two eigenvectors of $U$ are $|\pm \omega \ra = \frac{1}{\sqrt{2}}(|0\ra \pm i |1\ra)$ 
with eigenvalues $e^{\mp i \phi}$. The initial state is a uniform superposition of the two eigenvectors $|\Phi\ra=\frac{1}{\sqrt{2}} (|\omega\ra +|-\omega\ra)$. After $T$ applications of $U$, with $T$ chosen such that $T \phi=\frac{\pi}{2}$, we have 
$$U^T |\Phi\ra = U^T\frac{1}{\sqrt{2}}(|\omega\ra + |-\omega\ra)=\frac{1}{\sqrt{2}}(-i|\omega\ra +i|-\omega\ra)=|1\ra$$
which has an overlap of $\sqrt{\frac{N-1}{N}}$ with the marked state $|v\ra$.

In the random walk algorithm the transformation $(I-2\ket{s,v}\bra{s,v})$
is a counterpart of $R_v$ and the transformation $U$ is an ``imperfect'' 
counterpart of $R_{|\Phi\ra}$. 
We will first show, that with an appropriate choice of coin 
(as in Thms. \ref{Th:1}, \ref{Th:5}, and \ref{Th:3}) the resulting transformation 
is still {\em approximately} in a $2$-dimensional subspace; 
In this space $U'$ will correspond to a rotation as in Eq. (\ref{eq:rot}). 
Chosing $T$ appropriately will (approximately) give a state 
with a ``large'' overlap with the marked state or its neighbors. 

In the case of the ``bad'' coin, as in Theorem \ref{Th:2}, we will show that there is a large eigenspace of eigenvalue $1$ of the perturbed walk, and that the initial state has a large overlap with this eigenspace. Hence the state of the system nearly doesn't change by the walk.
\remove{
Note that for all walks defined here both $U$ and $U'=U(I-2 |s,v \ra \la s,v|)$ are real unitary
matrices. 
We will repeatedly use 

\begin{fact} \label{fact:1}
The eigenvectors of a real unitary matrix either have eigenvalue $\pm 1$
or else they appear in conjugated pairs with eigenvalues $e^{\pm i
\omega}$ and eigenvector $|\pm \omega \ra = \frac{1}{\sqrt{2}} (|R\ra
\pm i |I\ra)$, where $|R\ra$ and $|I\ra $ are real normalised vectors
and $\la R|I\ra=0$.
\end{fact}
{\em Proof:} trivial. 
}

More formally, an {\em abstract search algorithm} consists of two unitary 
transformations $U_1$ and $U_2$ and two states 
$\ket{\psi_{start}}$ and $\ket{\psi_{good}}$.
We require the following properties:
\begin{enumerate}
\item
$U_1=I-2\ket{\psi_{good}}\bra{\psi_{good}}$ (in other words, $U\ket{\psi_{good}}=-
\ket{\psi_{good}}$ and, if $\ket{\psi}$ is orthogonal to $\ket{\psi_{good}}$,
then $\ket{\psi}=\ket{\psi}$);
\item
$U_2\ket{\psi_{start}}=\ket{\psi_{start}}$ for some state $\ket{\psi_{start}}$ with 
real amplitudes and there is no other eigenvector with eigenvalue 1;
\item
$U_2$ is described by a real unitary matrix.
\end{enumerate}
The abstract search algorithm applies the unitary transformation
$(U_2 U_1)^T$ to the starting state $\ket{\psi_{start}}$.
We claim that, under certain constraints, its final state
$(U_2 U_1)^T\ket{\psi_{start}}$ has a sufficiently large  inner product with
$\ket{\psi_{good}}$.

The next lemmas, which we will prove in Sec. \ref{sec:prove}, describe the main properties of an abstract search algorithm
that we use. Let $U'=U_2 U_1$.
Since $U_2$ is a real unitary matrix, its non-$\pm 1$-eigenvalues come
in pairs of complex conjugate numbers. 
Denote them by $e^{\pm i \theta_1}$, $\ldots$, $e^{\pm i \theta_m}$. 
Let $\theta_{min}=\min(\theta_1, \ldots, \theta_m)$.


\begin{Lemma}\label{claim:1}
Define the arc $\cal{A}$ as the set of $e^{i\theta}$ for
all real $\theta$ satisfying $-\theta_{min}<\theta<\theta_{min}$.
Then $U'$ has at most two eigenvalues\footnote{The next lemma implies that there are
exactly two eigenvalues in $\cal{A}$.}
in $\cal{A}$.
\end{Lemma}

The two eigenvectors with these eigenvalues will be very important to us.
We will show that the starting state is close to a linear combination of them.
Therefore, we will be able to determine the evolution of the starting 
state by studying these two eigenvectors.

We start by bounding the two eigenvalues.
Let $\ket{\Phi_j^+}$ and $\ket{\Phi_j^-}$ be the eigenvectors with
eigenvalues $e^{i\theta_j}$ and $e^{-i\theta_j}$, respectively.
We express $\ket{\psi_{good}}$ as a superposition of
the eigenvectors of $U_2$:
\begin{equation} \label{eq:decomposition}
\ket{\psi_{good}}=a_{0}\ket{\psi_{start}}+\sum_{j=1}^m 
\left( a^+_j \ket{\Phi^+_j} + a^-_j \ket{\Phi^-_j} \right) .
\end{equation}

\begin{Lemma}
\label{lem:equal}
It is possible to select $\ket{\Phi_j^+}$ and $\ket{\Phi_j^-}$
so that $a^+_j=a^-_j$ and $a^+_j$ is a real number.
\end{Lemma}

In the next lemmas, we assume that this is the case and denote
$a_j^+=a_j^-$ simply as $a_j$.

\begin{Lemma}
\label{lem:1}
The eigenvalues of $U'$ in $\cal{A}$ are $e^{\pm i\alpha}$ where
\begin{equation}
\label{eq:sumlem1} 
\alpha=\Theta \left( \frac{1}{ \sqrt{\sum_j \frac{a^2_j}{a^2_0} 
\frac{1}{1-\cos \theta_{j}}} }\right) .
\end{equation}
\end{Lemma}

Let $\ket{w_{\alpha}}$ and $\ket{w_{-\alpha}}$ be the two eigenvectors with eigenvalues
$e^{i\alpha}$ and $e^{-i\alpha}$, respectively.
Define 
$\ket{w'_{start}}=\frac{1}{\sqrt{2}} \ket{w_{\alpha}}-
\frac{1}{\sqrt{2}} \ket{w_{-\alpha}}$, $\ket{w_{start}}=\frac{1}{\|w'_{start}\|} \ket{w'_{start}}$.
We claim that $\ket{w_{start}}$ is close to the starting state $\ket{\psi_{start}}$.
This is quantified by the following lemma.

\begin{Lemma}
\label{lem:2}
Assume that $\alpha<\frac{1}{2} \theta_{min}$. Then,
\[ \lbra \psi_{start} | w_{start}\rket \geq
1- \Theta\left( \alpha^4 
\sum_j \frac{a_j^2}{a_0^2} \frac{1}{(1-\cos\theta_{j})^2} \right) .\]
\end{Lemma}

The last lemma shows that, after repeating $U_2 U_1$ a certain number of times,
the state has significant overlap with $\ket{\psi_{good}}$.
Say we apply $(U_2 U_1)^{\lceil \pi/4\alpha \rceil}$ to the state 
$\frac{1}{\sqrt{2}} \ket{w_{\alpha}} - \frac{1}{\sqrt{2}}\ket{w_{-\alpha}}$.
Then, we get the state which is equal to
\[ \frac{1}{\sqrt{2}} e^{i\pi/4} \ket{w_{\alpha}} - e^{-i\pi/4} \frac{1}{\sqrt{2}}\ket{w_{-\alpha}}=
i(\frac{1}{\sqrt{2}} \ket{w_{\alpha}} + \frac{1}{\sqrt{2}}\ket{w_{-\alpha}})=:\ket{w_{good}} \]
plus a state of norm $O(\alpha)$ (because $\pi/4$ and  $\lceil \pi/4\alpha \rceil\alpha$ differ by an
amount which is less than $\alpha$).

\begin{Lemma}\label{lem:3}
Assume that $\alpha<\frac{1}{2} \theta_{min}$.
Let $\ket{w_{good}}=\frac{1}{\sqrt{2}} \ket{w_{\alpha}} +
\frac{1}{\sqrt{2}}\ket{w_{-\alpha}}$. 
Then, 
\[ |\lbra \psi_{good} | w_{good} \rket| = 
\Theta\left(\min\left( \frac{1}{\sqrt{\sum_j a_j^2 \cot^2 \frac{\theta_j}{4}}}\right), 
1\right). \]
\end{Lemma}

\begin{corollary}\label{cor:1}
Assume that $\alpha<\frac{1}{2} \theta_{min}$.
\[ |\lbra \psi_{good} | (U_2 U_1)^{\lceil \pi/4\alpha \rceil} | w_{good}\rket = 
\Theta\left(\min\left( \frac{1}{\sqrt{\sum_j a_j^2 \cot^2 \frac{\theta_j}{4}}}\right),
1\right) + O(\alpha) .\]
\end{corollary}

These three lemmas are the basis of our proofs. 
In each of our positive results, we first find a subspace $\H$ such that
the search algorithm restricted to this subspace is a special 
case of an abstract search algorithm.
Then, we apply Lemma \ref{lem:2} to show that the starting
state is close to $\ket{w_{start}}$ and Lemma \ref{lem:1} and corollary \ref{cor:1}
to show it evolves to a state having significant overlap with
$\ket{\psi_{good}}$.

\section{Proofs of the main results}

\subsection{Theorem 1}
\noindent Let us determine the eigenspectrum of $U=S_{ff} \cdot (C_0 \otimes I_N)$ first.

\begin{claim}{[Spectrum of $U$:]}\label{claim:spec}
$U$ has  eigenvalues $\lambda_{kl}$ with corresponding eigenvectors of the form
$|v_{kl}\ra \otimes |\chi_k\ra \otimes |\chi_l \ra$  for all $k,l=0, \ldots
,\sqrt{N}-1$, where $|\chi_k\ra = \frac{1}{\sqrt[4]{N}}
\sum_{j=0}^{\sqrt{N}-1}  \omega^{kj} |j\ra$ 
with $\omega=e^{2 \pi i / \sqrt{N}}$, and $\lambda_{kl}$ and $|v_{kl}\ra$
satisfy the equation
\begin{equation} \label{eq:ckl}
C_{kl} |v_{kl}\ra=\left( \begin{array}{cccc} 0 & \omega^{-k} & 0 & 0 \\
\omega^k & 0 & 0 & 0 \\ 0 & 0 & 0 & \omega^{-l} \\ 0 & 0 & \omega^{l} &
0 \end{array} \right)  \cdot C_0 |v_{kl}\ra = \lambda_{kl} |v_{kl}\ra.
\end{equation}
The four eigenvalues $\lambda_{kl}$ of $C_{kl}$ are $1, -1$ and $e^{\pm i \theta_{kl}}$
where $\cos \theta_{kl}=\frac{1}{2} (\cos \frac{2 \pi k}{\sqrt{N}} +
\cos \frac{2 \pi l}{\sqrt{N}} )$. 
Let $\ket{v^1_{kl}}$, $\ket{v^{-1}_{kl}}$ and $\ket{v^{\pm}_{kl}}$
be the vectors $\ket{v_{kl}}$ for the eigenvalues 1, -1 and $e^{i\pm \theta}$,
respectively.
Then, $|v_{kl}^{ 1}\ra$ is orthogonal to $|s\ra$ for $(k,l)\neq (0,0)$
and $|v_{kl}^{-1}\ra$ is orthogonal to $|s\ra$ for all $(k,l)$, including $(0,0)$. 
\end{claim}

\begin{proof}
Apply $U$ to a vector of the form $|v_{kl}\ra \otimes |\chi_k\ra \otimes
|\chi_l \ra$ and note that
$S_{ff}|\uparrow\ra \otimes |\chi_k\ra \otimes |\chi_l\ra =$ $
\omega^{- k} \ket{\downarrow} \otimes |\chi_k\ra \otimes |\chi_l\ra$, and
$S_{ff}|\downarrow\ra \otimes |\chi_k\ra \otimes |\chi_l\ra =
\omega^{k} |\uparrow\ra \otimes |\chi_k\ra \otimes |\chi_l\ra$, 
and similarly for the $y$-coordinate, which gives Eq. (\ref{eq:ckl}).  Solving the equation $|C_{kl}-\lambda I|=0$ for $\lambda$ gives the eigenvalues.
For $(k,l) \neq (0,0)$ the $1$-eigenvector $|v_{kl}^1\ra $ is proportional to
$ (\omega^k(\omega^l-1),1-\omega^l,\omega^l(1-\omega^k),\omega^k-1)$ and hence 
orthogonal to $|s\ra=\frac{1}{2}(1,1,1,1)$. The  
$-1$-eigenvector $|v_{kl}^{-1}\ra $ is proportional to
$ (\omega^l+1,\omega^k(\omega^l+1),-(\omega^k+1),-\omega^l(\omega^k+1))$ 
and hence 
orthogonal to $|s\ra=\frac{1}{2}(1,1,1,1)$. 
For $(k,l)=(0,0)$, $|s\ra$ is a $1$-eigenvector of $C_{00}$.
\end{proof}

For $(k,l)= (0,0)$, the eigenvalue 1 occurs 3 times. Thus, there is a 3-dimensional 
$1$-eigenspace. Since $\ket{s}$ is orthogonal to $\ket{v^{-1}_{00}}$,
$\ket{s}$ belongs to this eigenspace. 
We choose $|v_{00}^1\ra=|s\ra$ and $|v_{00}^\pm\ra$ orthogonal to $|s\ra$.

\remove{The ``gap'' $g$ between $1$ and the real part of the next eigenvalue $e^{i \theta_{01}}=e^{i \theta_{10}}$ on the unit circle is $g=1-Re (e^{i \theta_{01}})=1-\cos \theta_{01}=1-\frac{1}{2}(1+\cos\frac{2 \pi}{\sqrt{N}}) \approx \frac{1}{4}(\frac{2 \pi}{\sqrt{N}})^2=\frac{\pi^2}{N}$.
The eigenspace of eigenvalue $1$ of $U$ is $N$-fold degenerate and contains the initial state $|\Phi_0\ra$. The following shows that the $1$-eigenvectors of $U$ orthogonal to $|\Phi_0\ra$ are precisely the $1$-eigenvectors of $U'$. However, $U'$ has no transitions from $|\Phi_0\ra$ to $1$-eigenvectors of $U$ orthogonal to $|\Phi_0\ra$. 
}
 Let ${\cal H}'_0$ be the space spanned by the eigenvectors 
$\ket{v_{kl}^{\pm}}\otimes\ket{\chi_k}\otimes\ket{\chi_l}$, $(k, l)\neq (0, 0)$
and $|\Phi_0\ra=\ket{v_{00}^1}\otimes\ket{\chi_k}\otimes\ket{\chi_l}$.
Notice that all other eigenvectors of $U$ are orthogonal to $\ket{s, v}$,
by Claim \ref{claim:spec}.
Therefore, $\ket{s, v}$ is in ${\cal H}'_0$.
Moreover, applying $U'$ keeps the state in ${\cal H}'_0$, as shown by

\begin{claim}
\label{claim:0}
We have $U'({\cal H}'_0)={\cal H}'_0$.
Furthermore, $U'$ has no
eigenvector of eigenvalue $1$ in ${\cal H}'_0$.
\end{claim}

\begin{proof}
For the first part, notice that $U'=U(I-2|s, v\ra \la s, v|)$.
Therefore, it suffices to show $U({\cal H}'_0)={\cal H}'_0$
and $(I-2|s, v\ra \la s, v|)({\cal H}'_0)={\cal H}'_0$.
The first equality is true because ${\cal H}'_0$ has a basis consisting
of eigenvectors of $U$. Each of those eigenvectors gets mapped to
a multiple of itself which is in ${\cal H}'_0$. Therefore, 
$U({\cal H}'_0)={\cal H}'_0$.
The second equality follows because $(I-2|s, v\ra \la s, v|)|\psi\ra=
|\psi\ra-\la s, v|\psi\ra |s, v\ra$. This is a linear combination
of $|\psi\ra$ and $|s, v\ra$ and, if $|\psi\ra\in {\cal H}'_0$, it is in
${\cal H}_0$.

For the second part assume $|\omega_0\ra$ is an eigenvector of
eigenvalue $1$ of $U'$ in ${\cal H}'_0$. Then
$$0 \neq \la \Phi_0|\omega_0\ra=\la \Phi_0| U' |\omega_0\ra =\la \Phi_0| U (I-2|s,v\ra \la s,v|)|\omega_0\ra = \la
\Phi_0|\omega_0\ra -2 \la \Phi_0|s,v\ra \la s,v|\omega_0\ra.$$
This implies $ \la \Phi_0|s,v\ra \la s,v|\omega_0\ra=0$ and, since  $ \la \Phi_0|s,v\ra =\frac{1}{\sqrt{N}} \neq 0$, that
$\la s,v | \omega_0\ra=0$ and $|\omega_0\ra= U' | \omega_0\ra
=U|\omega_0\ra$ which in turn implies that $|\omega_0\ra$ is an eigenvector
of eigenvalue $1$ of $U$. Since $|\omega_0\ra$ has zero overlap with $|s,v\ra$ and precisely the $1$-eigenvectors of $U$ orthogonal to $|\Phi_0\ra$ have zero overlap with $|s,v\ra$, it follows that $\la \Phi_0|\omega_0\ra=0$ which contradicts that $|\omega_0\ra \in {\cal H}'_0$.
\end{proof}
The above shows that the random walk algorithm starting in $|\Phi_0\ra$ is restricted to a subspace ${\cal H}'_0$ of the
Hilbert space. 
Since $\ket{\Phi_0}$ is the only 1-eigenvector of $U$ in ${\cal H}'_0$,
we have an instance of the abstract search algorithm on the space 
${\cal H}'_0$,
with $U_1=I-2\ket{s,v}\bra{s,v}$, $U_2=U$, $\ket{\psi_{good}}=\ket{s, v}$
and $\ket{\psi_{start}}=\ket{\Phi_0}$.

As described in Section \ref{sec:abstract}, we study 
the 2-dimensional subspace spanned by $\ket{w_{\alpha}}$ and $\ket{w_{-\alpha}}$.
First, we bound $\alpha$ using Lemma \ref{lem:1}. 
We need to expand $|\psi_{good}\ra=|s,v\ra$ in the basis of eigenvectors of U.
Define $|\Phi^+_{kl}\ra=\ket{v_{kl}^{+}}\otimes\ket{\chi_k}\otimes\ket{\chi_l}$.
Let $\ket{\Phi^-_{kl}}$ be the vector obtained by replacing every amplitude
in $\ket{\Phi^+_{kl}}$ by its conjugate.
Then, $\ket{\Phi^-_{kl}}=\ket{v_{-k, -l}^{-}}\otimes\ket{\chi_{-k}}\otimes\ket{\chi_{-l}}$.
(This follows from two observations. First, replacing every amplitude
by its conjugate in $\ket{\chi_k}$ gives $\ket{\chi_{-k}}$. Therefore, 
$\ket{\Phi^-_{kl}}=\ket{v}\otimes\ket{\chi_{-k}}\otimes\ket{\chi_{-l}}$.
Second, since $U\ket{\Phi^+_{kl}}=e^{i\theta_{kl}} \ket{\Phi^+_{kl}}$,
we have $U\ket{\Phi^-_{kl}}=e^{-i\theta_{kl}} \ket{\Phi^-_{kl}}$,
implying that $\ket{v}=\ket{v^-_{-k,-l}}$.)
From Lemma \ref{lem:equal}, we have
\[ \ket{s, v}=a_{0} \ket{\Phi_0}+ 
\sum_{(k,l)\neq (0,0)} a_{kl} (\ket{\Phi_{kl}^+}+\ket{\Phi_{kl}^-} ) \]
where $|\Phi^+_{kl}\ra$ and $|\Phi^-_{kl}\ra$ appear with the same real coefficient 
$a_{kl}=\la s,v |\Phi^+_{kl}\ra=\la s,v |\Phi^-_{kl}\ra$.

\begin{claim}
\[ a_{kl} = \frac{1}{\sqrt{2N}} .\]
\end{claim}

\begin{proof}
We have $|\la v | \chi_k \ra \otimes |\chi_l \ra | =\frac{1}{\sqrt{N}}$
(since each of the $N$ locations has an equal emplitude in $| \chi_k \ra \otimes |\chi_l \ra$). 
It remains to show that $|\la s | v^+_{kl}\ra|=\frac{1}{\sqrt{2}}$.
For that, we first notice that $\ket{s}$ is a superposition of $\ket{v^{\pm}_{kl}}$
(since $\ket{v^{1}_{kl}}$ and $\ket{v^{-1}_{kl}}$ are orthogonal to $\ket{s}$).
By direct calculation $\la s | C_{kl} | s\ra = \frac{1}{2} (\cos \frac{2\pi k}{\sqrt{N}}+
\cos \frac{2\pi l}{\sqrt{N}})$.
We have 
\[ \la s | C_{kl} | s\ra = e^{i\theta_{kl}} \la s | v^+_{kl}\ra \la v^+_{kl} | s\ra +
e^{-i\theta_{kl}} \la s | v^-_{kl}\ra \la v^-_{kl} | s\ra .\]
This is possible only if $|\la s| v^+_{kl}\ra|=|\la s| v^-_{kl}\ra|=\frac{1}{\sqrt{2}}$.  
%
\end{proof}
Therefore,   
\begin{equation}
\label{eq-a10spec}
\ket{s, v}
=\frac{1}{\sqrt{N}} \ket{\Phi_0}+\frac{1}{\sqrt{2N}}
\sum_{(k,l)\neq (0,0)} (\ket{\Phi_{kl}^+}+\ket{\Phi_{kl}^-} ) .
\end{equation}
By Lemma \ref{lem:1}, $\alpha=\Theta(\frac{1}{\sqrt{\sum_{kl} \frac{1}{1-\cos \theta_{kl}}}})$.
The following claim implies
that $\alpha =\Theta( \frac{1}{\sqrt{c N\log N}})$.

\begin{claim}
\label{lem:thetasum}
$\sum_{(k,l) \neq (0,0)} \frac{1}{1-\cos\theta_{kl}} = \Theta(N\log N)$.
\end{claim}

\begin{proof}[ of Claim \ref{lem:thetasum}]

Recall from Claim \ref{claim:spec} that the eigenvalues corresponding to
$|v_{kl}^\pm\ra \otimes \ket{\chi_k}\otimes\ket{\chi_l}$
are $e^{\pm i\theta_{k l}}=\cos \theta_{kl}\pm i \sin \theta_{kl}$
where $\cos \theta_{kl}=\frac{1}{2}(\cos \frac{2 \pi k}{\sqrt{N}}+ \cos
\frac{2 \pi l}{\sqrt{N}})$.
For $x\in[0, 2\pi]$, we have
\begin{equation}
\label{eq-a3}
 1-\frac{x^2}{2}\leq \cos x \leq 1-\frac{2x^2}{\pi^2}.
\end{equation}
Therefore,
\[ 1-\frac{1}{4N} (k^2+l^2) \leq  \cos \theta_{kl} \leq 1-\frac{1}{\pi^2 N}
(k^2+l^2) ,\]
\begin{equation}
\label{eq-a4}
\frac{1}{\pi^2N} (k^2+l^2) \leq 1-\cos \theta_{kl} \leq \frac{1}{4N}
(k^2+l^2)
\end{equation}
This means that it suffices to show
\begin{equation}\label{eq:klsum}
N \sum_{k, l} \frac{1}{k^2+l^2} = \Theta(N \log N) ,
\end{equation}
where the summation is over all $k, l\in\{0, \ldots, \sqrt{N}-1\}$
such that at least one of $k, l$ is non-zero.
This follows because $\sum_{k, l}\frac{1}{k^2+l^2}=\Theta(\log N)$. A simple way to see this is to sum points that lie on  $m$-rectangles with the four corners $(\pm m, \pm m)$. The term $\frac{1}{k^2+l^2}$ for $(k,l)$ on an $m$-rectangle is bounded as $\frac{1}{2m^2} \leq \frac{1}{k^2+l^2} \leq \frac{1}{m^2}$, and there are $8m$ such points on each $m$-rectangle. Hence 
\begin{equation} \label{eq-a5}
\sum_{m=1}^{\sqrt{N}-1} 8m \frac{1}{2 m^2} \leq \sum_{k, l} \frac{1}{k^2+l^2} 
\leq\sum_{m=1}^{\sqrt{N}-1} 8m \frac{1}{ m^2}.
\end{equation}
The claim now follows from $\sum_{m=1}^{\sqrt{N}-1} \frac{1}{ m} = \frac{1}{2} \log N(1+o(1))$.
\end{proof}

Next, we use Lemma \ref{lem:2} to bound the overlap between $\ket{\Phi_0}$ and 
$\ket{w_{start}}=\frac{1}{\sqrt{2}} \ket{w_{\alpha}}-\frac{1}{\sqrt{2}} \ket{w_{-\alpha}}$.

\begin{claim}
\label{lem:thetasum1}
$\sum_{(k,l) \neq (0,0)} \frac{1}{(1-\cos\theta_{kl})^2} = \Theta(N^2)$.
\end{claim}

\begin{proof} [ of Claim \ref{lem:thetasum1} ]
Using the proof of Claim \ref{lem:thetasum}, 
\[ \frac{\pi^4 N^2}{(k^2+l^2)^2} \leq \frac{1}{(1-\cos \theta_{kl})^2} 
\leq \frac{4 N^2}{(k^2+l^2)^2} .\]
Therefore, it suffices to bound $N^2 \sum_{(k, l)\neq (0, 0)} \frac{1}{(k^2+l^2)^2}$.
Again, we sum points $(k, l)$ over rectangles with corners $(\pm m, \pm m)$.
Each rectangle has $8m$ points, each of which contributes 
a term of order $\frac{1}{m^4}$ to the sum. 
Since $\sum_{m} 8m \frac{1}{m^4}=\sum_m \frac{8}{m^3}$ is bounded by a constant,
the lemma follows.
\end{proof}

This means that the overlap between the starting state and $\ket{w_{start}}$ is
$1-\Theta(\alpha^4 N^2)=1-\Theta(\frac{1}{\log^2 N})$.
Equivalently, $\ket{\Phi_0}=\ket{w_{start}}+\ket{\Phi_{rem}}$, with
$\|\ket{\Phi_{rem}}\|=\Theta(\frac{1}{\log N})$.
After $\lceil \frac{\pi}{4\alpha} \rceil$ repetitions,
the state becomes $\ket{w_{good}}+\ket{\Phi'_{rem}}$,
with $\|\Phi'_{rem}\|= \Theta(\frac{1}{\log N})+O(\alpha)=\Theta(\frac{1}{\log N})$.
Finally, we bound $\lbra w_{good}|s, v\rket$, using Lemma \ref{lem:3}.
Since all $a_{k l}$ are equal to $\frac{1}{\sqrt{2N}}$ and $\cot x\leq \frac{1}{x}$, we have
\begin{equation}
\label{eq:lem3} 
\frac{1}{\sqrt{\sum_{k, l} a_{k l}^2 \cot^2 \frac{\theta_{k l}}{2}}} \geq 
\frac{1}{\sqrt{\frac{1}{2N}\sum_{k, l} \frac{4}{\theta_{k l}^2}}} .
\end{equation}
From the proof of Claim \ref{lem:thetasum}, we know that $\frac{1}{\theta_{k l}^2}$
is bounded from below and above by $\frac{const}{1-\cos\theta_{kl}}$.
Therefore, Claim \ref{lem:thetasum} implies
$\sum_{k, l} \frac{1}{\theta_{k l}^2} =\Theta(N \log N)$.
Thus, the expression of Eq. (\ref{eq:lem3}) is of order 
$\Omega(\frac{1}{\sqrt{\log N}})$. 

To conclude the proof of the theorem, the overlap of the state of the algorithm  after
$\lceil \frac{\pi}{4\alpha} \rceil$ steps and $\ket{s, v}$ is
\[ | \lbra w_{good}|s, v\rket +\lbra \Phi'_{rem} | s, v\rket | \geq 
|\lbra w_{good}|s, v\rket| - | \lbra \Phi'_{rem} | s, v\rket |.\]
The first term is of order $\Omega(\frac{1}{\sqrt{\log N}})$ and the second
term is of lower order ($\Omega(\frac{1}{\log N})$). Therefore,
the result is of order $\Omega(\frac{1}{\sqrt{\log N}})$.

Hence a measurement gives the marked location
$|v\ra$ with probability $p \geq \frac{c}{\log N}$.
This completes the proof of Theorem \ref{Th:1}.

\begin{proof}[ of Corollary 1] First, note that it is possible to generate the initial state $|\Phi_0\ra$ with $2 \sqrt{N}$ local transformations. We start with the state concentrated in one point (say $|0,0,0\ra$) and first ``spread'' the amplitude along the $x$-axis in $\sqrt{N}$ steps. In the first step we rotate the coin register to $\frac{1}{\sqrt[4]{N}}|0\ra + \sqrt{\frac{\sqrt{N}-1}{\sqrt{N}}}|1\ra$, followed by a $|1\ra$-controlled shift in the $x$-direction, followed by a rotation of the coin register back to $|0\ra$ in the vertex $(0,0)$. Similarly we repeat this procedure to move $\sqrt{\frac{\sqrt{N}-2}{\sqrt{N}}}$ of amplitude from $(1,0)$ to $(2,0)$ and so on. After $\sqrt{N}$ steps we have a uniform superposition over all vertices with $y$-coordinate  $0$. We repeat this process for the $y$-direction, which gives us the uniform superposition after another $\sqrt{N}$ steps. Note that this procedure also allows us to implement the reflection around the mean, $R_{|\Phi_0\ra}=I-2|\Phi_0\ra \la \Phi_0|$ in $4 \sqrt{N}$ steps: we simply run the procedure in reverse (which maps $|\Phi_0\ra$ to $|0,0,0\ra$), then invert the state $|0,0,0\ra$ (which can be done locally in the vertex $(0,0)$), and run the procedure forward again.

Note that we have determined the run-time $T$ only up to a constant 
(using Eqs. (\ref{eq-a4}) and  (\ref{eq-a5}),  (\ref{eq:upperalpha}), (\ref{eq-a2-2}) 
we can bound $T_{min}=\frac{\sqrt{N \log N}}{2} \leq T \leq \frac{\pi \sqrt{N \log N}}{2 \sqrt{2}}=T_{max}$). 
To get $\epsilon$-close to the state  $U'^T|\Phi_0\ra$ 
we use a standard trick and run the walk for times 
$T_{min},(1+\epsilon)T_{min},(1+\epsilon)^2T_{min},\ldots $ until we reach $T_{max}$. 
One of these times is within a factor of $(1 \pm \epsilon)$ of $T$ 
and hence our state and final measurement probability will be $\epsilon$-close 
to the state at time $T$. We can chose $\epsilon$ to be some small constant. 
The total time including all repetitions (bounded by $\frac{1}{1-\epsilon} T_{min}$) is still $O(T)=O(\sqrt{N \log N})$.

Finally, to amplify the success probability we will use amplitude amplification \cite{Brassard:02a}, which is a succession of steps consisting of reflection around the mean $|\Phi_0\ra$ and a run of the algorithm. The intermediate reflection around the mean can be implemented in $4 \sqrt{N}$ steps, the random walk takes $O(\sqrt{N \log N})$ steps, and we need $O(\sqrt{\log N})$ rounds of amplification to obtain a constant probability of success, which gives a total running time of $O(\sqrt{N} \log N)$.
\end{proof}

\subsection{Theorem 2}

\begin{proof}
The key difference between this walk using $S_m$ and the walk from Theorem
\ref{Th:1} using $S_{ff}$ is that the initial state $|\Phi_0\ra$ now has very large overlap with the eigenspace of eigenvalue $1$ of $U$ and $U'$. This
means that the walk (nearly) does not move at all and the state at any
time $T$ has overlap with $|\Phi_0\ra$ close to $1$. The difference becames apparent in the eigenspectrum of $U$:
\setcounter{claim}{5}
\begin{claim}\hskip -2.5mm ' {[Spectrum of U:]}\label{claim:spec1-1}
$U$ has  eigenvalues $\lambda_{kl}$ with corresponding eigenvectors of the form
$|v_{kl}\ra \otimes |\chi_k\ra \otimes |\chi_l \ra$  for all $k,l=0, \ldots
,\sqrt{N}-1$, where $|\chi_k\ra = \frac{1}{\sqrt[4]{N}}
\sum_{j=0}^{\sqrt{N}-1}  \omega^{kj} |j\ra$ 
with $\omega=e^{2 \pi i / \sqrt{N}}$, and $\lambda_{kl}$ and $|v_{kl}\ra$
satisfy the equation
\begin{equation} \label{eq:ckl1}
C_{kl} |v_{kl}\ra=\left( \begin{array}{cccc} \omega^{k} & 0 & 0 & 0 \\
0 & \omega^{-k} & 0 & 0 \\ 0 & 0 & \omega^{l} & 0 \\ 0 & 0 & 0 & 
\omega^{-l} \end{array} \right)  \cdot C_0 |v_{kl}\ra = \lambda_{kl} |v_{kl}\ra
\end{equation}
The four eigenvalues $\lambda_{kl}$ of $C_{kl}$ are $1, -1$ and $e^{\pm i \theta_{kl}}$,
where $\cos \theta_{kl}=-\frac{1}{2} (\cos \frac{2 \pi k}{\sqrt{N}} +
\cos \frac{2 \pi l}{\sqrt{N}} )$. 
For the eigenvector $|v_{kl}^{1}\ra$ corresponding
to eigenvalue $1$, we have 
\[ |\la v_{kl}^1 | s \ra|\geq \frac{1+\cos \frac{2 \pi k}{\sqrt{N}} +
\cos \frac{2 \pi l}{\sqrt{N}}+\cos \frac{2 \pi (k+l)}{\sqrt{N}}}{4} .\]
\end{claim}

\begin{proof}
The first part is by straightforward calculation as before (Claim \ref{claim:spec}).
For the second part, the eigenvector corresponding
to eigenvalue $1$ is $|v_{kl}^1\ra=\frac{|u_{kl}^1 \ra}{
\| u_{kl}^1 \|}$ with 
\[ |u_{kl}^1 \ra= ( w^k(1+w^l), 1+w^l, w^l(1+w^k), 1+w^k ) \]
We have $\|u_{kl}^1\| \leq 4$ because each of the $4$ components of 
$|u_{kl}^1\ra$ is at most $2$ in absolute value.
It remains to bound $\la u_{kl}^1 |s\ra$.
We have 
\[  \la v_{kl}^1|s\ra \geq \frac{\la u_{kl}^1 |s\ra}{4} = 
\frac{1}{8} (w^k(1+w^l)+1+w^l+w^l(1+w^k)+1+w^k)=
\frac{1}{4} (1+w^k)(1+w^l) .\]
The real part of this expression is 
$\frac{1}{4} (1+\cos \frac{2 \pi k}{\sqrt{N}}+\cos \frac{2 \pi l}{\sqrt{N}}
+\cos \frac{2 \pi (k+l)}{\sqrt{N}})$.
This implies the claim.
\end{proof}
Let $\H_1$ be the 1-eigenspace of $U$, spanned by the $|\Phi_{kl}^1\ra=|v_{kl}^1\ra \otimes |\chi_{k}\ra \otimes |\chi_{l}\ra$ for $k,l=0, \ldots , \sqrt{N}-1$, with $|\Phi_{00}^1\ra = |\Phi_0\ra$. Write
\[ \ket{s, v}=\sum_{k,l} \alpha_{kl} \ket{\Phi_{kl}^1} + \ket{s'} \]
where $\ket{s'}$ has no overlap with the $1$-eigenspace $\H_1$.
We claim
\setcounter{claim}{10}
\begin{claim}
$U'$ has a $1$-eigenvector $\ket{\Phi^\perp}$ such that
$|\lbra \Phi_0 | \Phi^\perp \rket|^2 =
1-\frac{|\alpha_{00}|^2}{\sum_{i, j=1}^{\sqrt{N}} |\alpha_{i j}|^2}$.
\end{claim}

\begin{proof}
Let $\beta_{kl}=\frac{\alpha_{kl}}{\sqrt{\sum_{i, j=1}^{\sqrt{N}} |\alpha_{i j}|^2}}$.
Let
$\ket{\Phi}=\sum_{k,l} \beta_{kl} \ket{\Phi_{kl}^1}$
be the projection of $\ket{s, v}$ on $\H_1$.
Since $\lbra \Phi_0|\Phi \rket=\beta_{00}$, we can write
\[ \ket{\Phi_0}=\beta_{00} \ket{\Phi}+ \sqrt{1-|\beta_{00}|^2} \ket{\Phi^{\perp}} \]
where $\ket{\Phi^{\perp}}$ is a vector perpendicular to $\ket{\Phi}$.
Since $\ket{\Phi_0}$ and $\ket{\Phi}$ are both in the subspace $\H_1$,
$\ket{\Phi^{\perp}}$ is also in $\H_1$.
We claim that $\ket{\Phi^{\perp}}$ is a $1$-eigenvector of $U'$.
The state $\ket{\Phi^{\perp}}$ is orthogonal to $\ket{s, v}$
because $\ket{\Phi^{\perp}}$ belongs to  $\H_1$ and
is orthogonal to $\ket{\Phi}$ which is the projection of $\ket{s, v}$ to
that subspace. Therefore, $\ket{\Phi^{\perp}}$ is a $1$-eigenvector of
$I-\ket{s, v}\bra{s, v}$. $\ket{\Phi^{\perp}}$ is also
a $1$-eigenvector of $U$ because it belongs to $\H_1$.
This means that it is a $1$-eigenvector of $U'=U(I-\ket{s, v}\bra{s, v})$ as
well.
\end{proof}
To complete the proof, we need to bound $|\alpha_{00}|$, $|\alpha_{01}|$,  $\ldots$, $|\alpha_{\sqrt{N}-1,\sqrt{N}-1}|$.
We have $\alpha_{00}=\frac{1}{\sqrt{N}}$.
We will show that there are $\Omega(N)$
other $\alpha_{kl}$ of order $\Omega(1/\sqrt{N})$. This would
imply that the overlap of $|\Phi_0\ra$ with a $1$-eigenvector of $U'$ is
\[
1-\frac{\alpha_{00}^2}{\sum_{i, j=1}^{\sqrt{N}} |\alpha_{i j}|^2}=1-\Omega(\frac{1}{N}) .\]
Claim \ref{claim:spec1-1}' gives the desired a bound on the $\alpha_{kl}$: 
\[ |\la s, v|v^1_{kl}\ra \otimes |\chi_k \ra \otimes |\chi_l\ra|  
 = |\la v | \chi_k \otimes \chi_l \ra| \times  |\la s | v^1_{kl} \ra| 
 = \frac{ |\la s | v^1_{kl} \ra| }{\sqrt{N}}\geq 
\frac{1+\cos \frac{2 \pi k}{\sqrt{N}}+\cos \frac{2 \pi l}{\sqrt{N}}+
\cos\frac{2 \pi (k+l)}{\sqrt{N}}}{4\sqrt{N}}.\]
The range for $\frac{2 \pi k}{\sqrt{N}}$ and $\frac{2 \pi l}{\sqrt{N}}$ is $[-\frac{\pi}{2}, \frac{\pi}{2}]$. Therefore, for half of all $k$ (resp. half of all $l$) we have   $|k|\frac{2 \pi}{\sqrt{N}}\leq \frac{\pi}{4}$ (resp.  $|l|\frac{2 \pi}{\sqrt{N}}\leq \frac{\pi}{4}$). For the $\frac{N}{4}$ pairs $(k, l)$ that satisfy both of those conditions we have
\[ 1+\cos \frac{2 \pi k}{\sqrt{N}}+\cos \frac{2 \pi l}{\sqrt{N}}+\cos\frac{2 \pi (k+l)}{\sqrt{N}} \geq 1+\frac{1}{\sqrt{2}}+\frac{1}{\sqrt{2}}\geq
1+\sqrt{2} .\]
Thus, for at least  $\frac{N}{4}$
pairs $(k, l)$
\[ |\alpha_{kl}|=|\la s, v|v^1_{kl}\ra \otimes |\chi_k\ra \otimes |\chi_l\ra| \geq
\frac{1+\sqrt{2}}{4\sqrt{N}} > \frac{1}{2\sqrt{N}}. \]
\end{proof}

\subsection{Theorem 3}
\begin{proof}
The proof for this random walk algorithm with a $2$-dimensional coin proceeds in close analogy to the proof of Theorem \ref{Th:1}, and we will emphasize and prove the points that differ.
\setcounter{claim}{5}
\begin{claim}\hskip -2.0mm {\bf ''}{[Spectrum of U:]}\label{claim:spec1-2}
$U$ has  eigenvalues $\lambda_{kl}^\pm$ with corresponding eigenvectors of the form
$|v_{kl}^\pm\ra \otimes |\chi_k\ra \otimes |\chi_l \ra$  for all $k,l=0, \ldots
,\sqrt{N}-1$, where $|\chi_k\ra = \frac{1}{\sqrt[4]{N}}
\sum_{j=0}^{\sqrt{N}-1}  \omega^{kj} |j\ra$ 
with $\omega=e^{2 \pi i / \sqrt{N}}$, and $\lambda_{kl}^\pm$ and $|v_{kl}^\pm\ra$
satisfy the equation
\begin{equation} \label{eq:ckl2}
C_{kl} |v_{kl}\ra=\left( \begin{array}{cc} 
\omega^l \cos k & i \omega^{-l} \sin k \\
i \omega^l \sin k & \omega^{-l} \cos k  
\end{array} \right)  |v_{kl}\ra = \lambda_{kl} |v_{kl}\ra.
\end{equation}
The two eigenvalues $\lambda_{kl}^\pm$ of $C_{kl}$ are $e^{\pm i \theta_{kl}}$
where $\cos \theta_{kl}=\frac{1}{2} (\cos \frac{2 \pi (k+l)}{\sqrt{N}} +
\cos \frac{2 \pi (k-l)}{\sqrt{N}} )$. 
\end{claim}

As a corollary, we have that there are exactly two eigenvectors 
with eigenvalue $1$, both of them of the form 
$|v_{00}^\pm\ra \otimes |\chi_0\ra \otimes |\chi_0 \ra$.
Since the coin space is $2$-dimensional, the two vectors 
$|v_{00}\ra$ span it and, therefore, 
$|v\ra \otimes |\chi_0\ra \otimes |\chi_0\ra$ is an
eigenvector for any $|v\ra$. 
In particular, we can take
$|v_{00}^+\ra=|s\ra$ and $|v_{00}^-\ra=|s^{\perp}\ra$
where $|s^{\perp}\ra \perp |s\ra$.
Similarly to Theorem 1, let $\H'_0$ be the space 
$|v_{00}^+\ra \otimes |\chi_0\ra \otimes |\chi_0 \ra$
and $|v_{kl}^\pm\ra \otimes |\chi_k\ra \otimes |\chi_l \ra$.
Similarly to Claim \ref{claim:0}, $\H'_0$ is mapped to itself by $U'$.

Let $\ket{\Phi_{kl}^+}=\ket{v_{kl}^+}\otimes \ket{\chi_k}\otimes \ket{\chi_l}$ and 
$\ket{\Phi_{kl}^-}=\ket{v_{-k, -l}^-}\otimes \ket{\chi_{-k}}\otimes \ket{\chi_{-l}}$.
We can express the state $\ket{s, v}$ as 
\begin{equation}
\label{eq-a10a}
\ket{s, v}=a_0 \ket{\Phi_0}+
\sum_{(k,l)\neq (0,0)} a_{kl} (\ket{\Phi_{kl}^+}+\ket{\Phi_{kl}^-} ).
\end{equation}
where the equality of the coefficients of $\ket{\Phi_{kl}^+}$ and $\ket{\Phi_{kl}^-}$
follows from the proof of Lemma \ref{lem:equal}
(and $\ket{\Phi_{kl}^+}$ and $\ket{\Phi_{kl}^-}$ being complex conjugates).


We now have to bound the sum of Eq. (\ref{eq:sumlem1}). 
We claim that replacing all $\frac{|a_{kl}|^2}{|a_0|^2}$ by $\frac{1}{2}$ does
not change the value of the sum.
To see that, we first notice that $\theta_{kl}=\theta_{-k,-l}$.
Therefore, replacing $|a_{kl}^2|$ and $|a_{-k,-l}^2|$ by 
$\frac{|a_{kl}|^2+|a_{-k,-l}|^2}{2}$ does not change the sum.
Futhermore, $|a_{kl}|^2+|a_{-k,-l}|^2=\frac{1}{N}$.
(We have $|a_{kl}|=\la \Phi^+_{kl}|s,v\ra$, 
$|a_{-k,-l}|=\la \Phi^+_{-k,-l}|s,v\ra=\la\Phi^-_{-k,-l}|s,v\ra$.
The vectors $\ket{\Phi^+_{kl}}$ and $\ket{\Phi^-_{-k,-l}}$ 
are the same as $\ket{v^{\pm}_{kl}}\otimes\ket{\chi_k}\otimes\ket{\chi_l}$.
Therefore, $|a_{kl}|^2+|a_{-k,-l}|^2$ equals the squared projection 
of $\ket{s, v}$ to $\ket{v^{\pm}_{kl}}\otimes\ket{\chi_k}\otimes\ket{\chi_l}$
which is equal to $|\la v | \chi_k \ra \otimes |\chi_l\ra|^2=\frac{1}{N}$.)
Also, we still have $a_0=\frac{1}{\sqrt{N}}$ and $|a_0|^2=\frac{1}{N}$.
Therefore, Eq. (\ref{eq:sumlem1}) simplifies to 
\[ \alpha=\Theta \left(\frac{1}{\sqrt{\sum_{(k,l)\neq (0,0)}  
\frac{1}{2(1-\cos\theta_{kl})} }}\right) ,\]
just as in the proof of Theorem \ref{Th:1}.
We get $\alpha=\Theta(\sqrt{N\log N})$ as in Lemma \ref{lem:thetasum}.
The other two parts of Theorem \ref{Th:1} also follow in a similar way.
\end{proof}
Similarly to Corollary 1 we can get a random walk based search algorith that determines the marked state with constant probability in time $O(\sqrt{N}\log N)$.

\subsection{Theorem 4}

\begin{proof}
Let us call the $2d$ directions on the $d$-dimensional grid $i_{\pm}$,
where $i$ indicates the dimension and $\pm$ the direction of the walk.
Then the ``flip-flop'' shift operation takes the form
\[S_{ff}| i_\pm \ra \otimes |x_1 \ldots x_i \ldots x_d\ra=|i_\mp\ra \otimes |x_1 \ldots x_i \pm 1 \ldots x_d \ra \]
Let us recapitulate what the key elements in the proof of Theorem
\ref{Th:1} are and how they generalize to the $d$-dimensional case.

\setcounter{claim}{5}
\begin{claim} \hskip -2.5mm {\bf{$^{\prime \prime \prime}$}} {[Spectrum of U:]}
 U has eigenvalues $\lambda_{k_1k_2 \ldots k_d}$ with corresponding eigenvectors
of the form $|v_{k_1k_2 \ldots k_d}\ra \otimes
|\chi_{k_1}\ra \otimes |\chi_{k_2} \ra \otimes \ldots \otimes
|\chi_{k_d}\ra$ ($k_i=0, \ldots ,\sqrt[d]{N}-1$), where $|\chi_i\ra =
\frac{1}{\sqrt[2d]{N}} \sum_{k=0}^{\sqrt[d]{N}-1}  \omega^k |k\ra$ with $\omega=e^{2 \pi i /
\sqrt[d]{N}}$, and $\lambda_{k_1k_2 \ldots k_d}$ and $|v_{k_1k_2 \ldots k_d}\ra$ satisfy the equation
\begin{equation} \label{eq:cki}
C_{k_1k_2 \ldots k_d} |v_{k_1k_2 \ldots k_d}\ra =  \left(
\begin{array}{ccccc} 0 & \omega^{-k_1} & 0 & 0 & \\ \omega^{k_1} & 0 & 0
& 0& \\ 0 & 0 & 0 & \omega^{-k_2} &...\\ 0 & 0 & \omega^{k_2} & 0
&\\&&...&& \end{array} \right)  \cdot C_0 |v_{k_1 k_2 \ldots k_d}\ra =
\lambda_{k_1k_2 \ldots k_d} |v_{k_1k_2 \ldots k_d}\ra.
\end{equation}
The $2d$ eigenvalues $\lambda_{k_1k_2 \ldots k_d}$ of $C_{k_1k_2 \ldots k_d}$ are $1$ and $-1$ with
multiplicity $d-1$ each, and $e^{\pm i \theta_{k_1k_2 \ldots k_d}}$ where
$\cos \theta_{k_1k_2 \ldots k_d}=\frac{1}{d} \sum_{i=1}^d \cos 
\frac{2 \pi k_i}{\sqrt[d]{N}} $. All $|v_{k_1k_2 \ldots k_d}^{1}\ra$
corresponding to eigenvalue $ 1$ are orthogonal to $|s\ra$ for
$(k_1k_2 \ldots k_d)\neq (0,0,\ldots,0)$. All $|v_{k_1k_2 \ldots k_d}^{-1}\ra$
corresponding to eigenvalue $-1$ are orthogonal to $|s\ra$ for
all $(k_1k_2 \ldots k_d)$.
\end{claim}

\begin{proof}
Eq. (\ref{eq:cki}) is obtained in the same way as in the proof of Claim
\ref{claim:spec}. $C_{k_1k_2 \ldots k_d}$ consists of 2 parts, the
2-block-diagonal matrix (call it $D_{k_1k_2 \ldots k_d}$) and $C_0$.
Each block in $D_{k_1k_2 \ldots k_d}$ has eigenvalues $\pm 1$ and
eigenvectors $(\omega^{-\frac{k_i}{2}},\pm \omega^{\frac{k_i}{2}})$, so
the matrix $D_{k_1k_2 \ldots k_d}$ itself has eigenspaces of $1$ and
$-1$ of dimension $d$ each. In each of these two eigenspaces we can find
 $d-1$ orthogonal vectors orthogonal to $|s\ra$. For those vectors
$C_0=2|s\ra \la s|-I$ just flips their sign, so the $-1$ eigenvectors of
 $D_{k_1k_2 \ldots k_d}$ become $+1$ eigenvectors of $C_{k_1k_2 \ldots
k_d}$ and vice versa. Call the remaining two eigenvectors of eigenvalue
$\pm 1$ $|e_\pm\ra$ and expand $|s\ra = \alpha_+ |e_+\ra + \alpha_-
|e_-\ra$. Then $\la s | C_{k_1k_2 \ldots k_d}|s\ra = \la s| D_{k_1k_2
\ldots k_d}|s\ra=\frac{1}{d} \sum_{i=1}^d \cos 2 \pi
k_i/\sqrt[d]{N}=\cos \theta_{k_1k_2 \ldots k_d}$ implies
$|\alpha_+|^2-|\alpha_-|^2=\cos \theta_{k_1k_2 \ldots k_d}$. Let
$|\omega\ra$ be an eigenvector of $C_{k_1k_2 \ldots k_d}$ not orthogonal
to $|s\ra$ and expand $|\omega\ra = \beta_+ |e_+\ra + \beta_- |e_-\ra$.
Then we obtain (using $|e_+\ra =\alpha_+^* |s\ra + \alpha_- |s^\perp
\ra$ and $|e_-\ra =\alpha_-^* |s\ra - \alpha_+ |s^\perp \ra$ where
$|s^\perp \ra$ is orthogonal to $|s\ra$ in the space spanned by
$|e_\pm\ra$) for the eigenvalue $\la \omega| C_{k_1k_2 \ldots
k_d}|\omega\ra=(|\alpha_+|^2-|\alpha_-|^2)+4 i Im \beta_+^*
\beta_-\alpha_+ \alpha_-^*=\cos \theta_{k_1k_2 \ldots k_d} \pm i \sin
\theta_{k_1k_2 \ldots k_d}$.
\end{proof}

For $(k_1k_2 \ldots k_d)= (0,0,\ldots,0)$ there are $d+1$ $1$-eigenvectors, we set $|v_{0 \ldots 0}\ra = |s\ra$.
Then, the other $1$-eigenvectors are orthogonal to $|s\ra$.

Similarly to Theorem \ref{Th:1}, we restrict to the subspace $\H'_0$ spanned by 
$\ket{v_{k_1 k_2\ldots k_d}^{\pm 1}}\otimes|\chi_{k_1}\ra \otimes |\chi_{k_2} \ra 
\otimes \ldots \otimes |\chi_{k_d}\ra$
and $\ket{v_{00\ldots 0}}\otimes|\chi_{0}\ra \otimes |\chi_{0} \ra 
\otimes \ldots \otimes |\chi_{0}\ra$. As in Theorem \ref{Th:1}, we have
$U'(\H'_0)=H'_0$. Further
\[ \ket{s, v}=\frac{1}{\sqrt{N}} \ket{\Phi_0}+\frac{1}{\sqrt{2N}}
\sum_{(k_1, k_2, \ldots,k_d)\neq (0, 0, \ldots, 0)} 
(\ket{\Phi_{k_1 k_2\ldots k_d}^+}+\ket{\Phi_{k_1 k_2\ldots k_d}^-} ) ,
\]
where $\ket{\Phi^{+}_{k_1 k_2\ldots k_d}}=
\ket{v_{k_1 k_2\ldots k_d}^{+ 1}}\otimes|\chi_{k_1}\ra \otimes |\chi_{k_2} \ra 
\otimes \ldots \otimes |\chi_{k_d}\ra$ and
$\ket{\Phi^{-}_{k_1 k_2\ldots k_d}}=
\ket{v_{k_1 k_2\ldots k_d}^{-1}}\otimes|\chi_{-k_1}\ra \otimes |\chi_{-k_2} \ra 
\otimes \ldots \otimes |\chi_{-k_d}\ra$.
We use a modified Claim \ref{lem:thetasum}:
\setcounter{claim}{8}
\begin{claim}\hskip -2.5mm ' 
$\sum_{(k_1,k_2, \ldots k_d) \neq (0,0,\ldots ,0)}
\frac{1}{1-\cos \theta_{k_1k_2 \ldots k_d}} = \Theta(N)$.
\end{claim}
\begin{proof}
The proof follows along the lines of the proof of Claim \ref{lem:thetasum}.
By Claim \ref{claim:spec}$^{\prime \prime \prime}$,
\[ \frac{1}{1-\cos \theta_{k_1k_2 \ldots k_d}} = 
\frac{d}{\sum_{j=1}^d (1-\cos \frac{2\pi k_j}{N^{1/d}})} \]
Similarly to Lemma \ref{lem:thetasum}, this is bounded from above and 
below by a constant times $N^{2/d}\frac{1}{k_1^2+k_2^2+\ldots+k_d^2}$.
Thus, we have to estimate
\begin{equation}
\label{eq:ddimsum}
N^{2/d} \sum_{k_1,k_2,\ldots, k_d} \frac{1}{k_1^2+k_2^2+ \ldots + k_d^2} 
\end{equation}
where the summation is over all $k_i\in\{0, \ldots, \sqrt[d]{N}-1\}$
such that at least one of the $k_i$ is non-zero.
We divide tuples the $(k_1, \ldots, k_d)$ into $N^{1/d}$ "slices",
with the $m^{\rm th}$ "slice" containing those tuples
where $\max (k_1, k_2, \ldots, k_d)=m$. 
The $m^{\rm th}$ slice contains $O(m^{d-1})$ tuples.
Therefore, the sum $\sum_{k_1, \ldots, k_d}\frac{1}{k_1^2+k_2^2+ \ldots + k_d^2}$
over the $m^{\rm th}$ slice is of order $m^{d-1}\frac{1}{m^2}=m^{d-3}$.
Since there are $N^{1/d}$ slices and, for each of them, the sum is
$m^{d-3}\leq N^{(d-3)/d}$, the sum $\sum_{k_1, \ldots, k_d}\frac{1}{k_1^2+k_2^2+ \ldots + k_d^2}$
over all $(k_1, \ldots, k_d)\neq (0, \ldots, 0)$ is of order at most
$N^{1/d} N^{(d-3)/d}=N^{(d-2)/d}$.
This implies that Eq. (\ref{eq:ddimsum}) is of order at most $N$.
It is also of order at least $N$ since each individual term
$N^{2/d}\frac{1}{k_1^2+k_2^2+\ldots+k_d^2}$ is at least a constant.
\end{proof}
Therefore, applying Lemma \ref{lem:1} gives a sharper bound 
$\alpha=\Theta(\frac{1}{\sqrt{N}})$.
Notice that we still fulfill the requirement $\alpha<\frac{1}{2}\theta_{min}$ 
 needed 
for Lemmas \ref{lem:2} and \ref{lem:3}. The reason for that is 
that all $\theta_i$ are at least $\Omega(\frac{1}{N^{1/d}})$.

Similarly to Claim \ref{lem:thetasum}', we can show
\[ \sum_{(k_1,k_2, \ldots k_d) \neq (0,0,\ldots ,0)}
\frac{1}{(1-\cos \theta_{k_1k_2 \ldots k_d})^2} = \Theta(N) \]
and
\[ \sum _{(k_1,k_2, \ldots k_d) \neq (0,0,\ldots ,0)}
\frac{1}{\cot \theta_{k_1k_2 \ldots k_d}^2} = \Theta(N) .\]
By combining these two equalities with
Lemmas \ref{lem:2} and \ref{lem:3}, we get that the overlap
between the starting state $\ket{\Phi_0}$ and $|w_{start}\ra=\frac{1}{\sqrt{2}}\ket{w_{\alpha}}-
\frac{1}{\sqrt{2}}\ket{w_{-\alpha}}$ is $1-\Theta(\frac{1}{\sqrt{N}})$ and
the overlap between $|w_{good}\ra=\frac{1}{\sqrt{2}}\ket{w_{\alpha}}+\frac{1}{\sqrt{2}}\ket{w_{-\alpha}}$
and $\ket{s, v}$ is $\Omega(1)$.
This implies that the search algorithm's final state has a constant
overlap with $\ket{s, v}$.
\end{proof}

\subsection{Theorem 5}

We first discuss generalizing the positive results (Theorems \ref{Th:1}, 
\ref{Th:3} and \ref{Th:4}) to the case with two marked items.
The main issue is to state them as instances of the abstract search.
Assume there are $k$ marked locations $v_1, \ldots, v_k$.
Then, one step of the search algorithm is $U'=(I-2\sum_{i=1}^k \ket{s, v_i}\bra{s, v_i}) U$.
Currently, we are not able to analyze cases of the abstract search where $U_2$ flips the 
sign on more than a 1-dimensional subspace.

For the $k=2$ case, we can avoid this problem, via a reduction to $k=1$.
Define $\ket{s'}=\frac{1}{\sqrt{2}}\ket{s, v_1}+\frac{1}{\sqrt{2}}\ket{s, v_2}$.
We claim that applying $(U')^T$ to the starting state $\ket{\Phi_0}$
gives the same final state as applying $(U'')^T$ where $U''=(I-2\ket{s'}\bra{s'})U$.

To show that, let $T$ be a symmetry of the grid such that $T(v_1)=v_2$ and $T(v_2)=v_1$.
(For the 2-dimensional grid, if $v_1=(x_1, y_1)$ and $v_2=(x_2, y_2)$, then
$T(x, y)=(x_1+x_2-x, y_1+y_2-y)$.)
We identify $T$ with the unitary mapping $\ket{c, v}$ to $\ket{c, T(v)}$.

\setcounter{claim}{11}
\begin{claim}
For any $t\geq 0$, $T (U')^t\ket{\Phi_0} =(U')^t \ket{\Phi_0}$.
\end{claim}

\begin{proof}
By induction. For the base case, we have to show $T\ket{\Phi_0}=\ket{\Phi_0}$.
This follows since $\ket{\Phi_0}$ is a uniform superposition of the states
$\ket{s, v}$ and $T$ just permutes locations $v$.

For the inductive case, notice that $T$ commutes with both $I-2\ket{s, v_1}\bra{s, v_1}-
2\ket{s, v_2}\bra{s, v_2}$ and $U$. Therefore, $TU'=U'T$.
If the inductive assumption $T (U')^t\ket{\Phi_0} =(U')^t \ket{\Phi_0}$ is true, then
we also have
\[ T (U')^{t+1}\ket{\Phi_0} =U' T (U')^t \ket{\Phi_0} = (U')^{t+1} \ket{\Phi_0} ,\]
completing the induction step. 
\end{proof}

Let $\ket{s''}=\frac{1}{\sqrt{2}}\ket{s, v_1}-\frac{1}{\sqrt{2}}\ket{s, v_2}$.
Then, $(U')^t \ket{\Phi_0}$ is orthogonal to $\ket{s''}$ because $T\ket{s''}=-\ket{s''}$.
We have $\ket{s, v_1}\bra{s, v_1}+\ket{s, v_2}\bra{s, v_2}=\ket{s'}\bra{s'}+\ket{s''}\bra{s''}$.
Together with $(U')^t \ket{\Phi_0} \perp \ket{s''}$, this means 
\[ (I-2\ket{s, v_1}\bra{s, v_1}-2\ket{s, v_2}\bra{s, v_2}) (U')^t \ket{\Phi_0} = 
(I-2\ket{s'}\bra{s'}) (U')^t \ket{\Phi_0} .\]
Thus, $U'$ can be replaced by $U''$ at every step.

The rest of the proofs is now similar to the case of 1 marked location, except that $\ket{s, v}$ is 
replaced by $\ket{s'}$ everywhere.
Theorem \ref{Th:2} also follows similarly to the 1 marked item case, 
with $\ket{s'}$ instead of $\ket{s, v}$.

\section{Proofs of the technical lemmas}\label{sec:prove}

In this section, we prove Lemmas \ref{claim:1}, \ref{lem:1}, 
\ref{lem:equal}, \ref{lem:2} and \ref{lem:3}.
We will repeatedly use the following result which can be found in many
linear algebra textbooks.

\begin{fact} \label{fact:1}
The eigenvectors of a real unitary matrix either have eigenvalue $\pm 1$
or else they appear in conjugated pairs with eigenvalues $e^{\pm i
\omega}$ and eigenvector $|\pm \omega \ra = \frac{1}{\sqrt{2}} (|R\ra
\pm i |I\ra)$, where $|R\ra$ and $|I\ra $ are real normalised vectors
and $\la R|I\ra=0$.
\end{fact}

\begin{proof} [ of Lemma \ref{claim:1}]
Let $g=1-Re (e^{i\theta_{min}})$. 
Then, for any $e^{i \theta} \in \cal{A}$, $Re (e^{i \theta}) >1-g$ and, for every eigenvector $|\omega\ra$ with
an eigenvalue in $\cal{A}$, $Re \la \omega | U' | \omega\ra > 1-g$.

If there were more than two eigenvectors of $U'$ in ${\cal H}'_0$ with eigenvalues on the arc
$\cal A$, we could construct a linear combination $|a\ra$ of them such
that $|a\ra \perp |\Phi_0\ra, |s,v\ra$. Since $|a\ra$ is a linear
combination of vectors $|\omega \ra$ with $Re \la \omega |U'|\omega\ra >
1-g$ we have $Re \la a|U'|a\ra >1-g$. But then $Re \la a|U'|a\ra =Re \la
a |U(I-2|s,v\ra \la s,v|)|a \ra = Re \la a|U|a\ra$. 
Since $|a\ra$ is orthogonal to $|\Phi_0\ra$ and all other $1$-eigenvectors of $U$ ($|a\ra \in {\cal H}'_0$), $|a\ra$ is a linear combination of eigenvectors of $U$ at least $g$ away from $1$ and hence $Re \la a|U|a\ra \leq 1-g$, which gives a
contradiction.
%
\end{proof}

\begin{proof}[ of Lemma \ref{lem:equal}] 
Let $\ket{\Phi}$ be the vector obtained by replacing every 
amplitude in $\ket{\Phi^+_j}$ by its complex conjugate. 
Since $U_2$ is a real unitary matrix, $U_2\ket{\Phi^+_j}=e^{i\theta_j}\ket{\Phi^+_j}$
implies $U_2\ket{\Phi}=e^{-i\theta_j}\ket{\Phi}$.
Therefore, we can assume that $\ket{\Phi^-_j}$ is
a complex conjugate of $\ket{\Phi^+_j}$.
The coefficients $a^+_j$ and $a^-_j$ are equal to the inner products
$\lbra \Phi_j^+ | \psi_{start}\rket$ and $\lbra \Phi_j^- | \psi_{start}\rket$. 
Since $\ket{\psi_{start}}$ is a real vector,
these two inner products are complex conjugates
and $a^+_j=(a^-_j)^*$.
By multiplying $\ket{\Phi^+_j}$ and $\ket{\Phi^-_j}$
with appropriate constants, we can achieve $a_j^+=a_j^-$.
\end{proof}

\begin{proof} [ of Lemma \ref{lem:1}]
First, we express $\ket{\psi_{good}}$ in the basis consisting of eigenvectors of $U_2$:
\begin{equation}
\label{eq-psigood}
\ket{\psi_{good}} = a_0 \ket{\Phi_0} + \sum_{j=1}^m a_j (\ket{\Phi_j^+}+\ket{\Phi_j^-}) .
\end{equation}
where $\ket{\Phi_0}=\ket{\psi_{start}}$.
We define for real $\alpha$
\begin{equation}
\label{eq-a11}
\ket{w'_{\alpha}}= a_0 \cot \frac{\alpha}{2} \ket{\Phi_0}
+ \sum_{j} a_j\left(
\cot \frac{\alpha-\theta_{j}}{2} \ket{\Phi_{j}^+}
+\cot \frac{\alpha+\theta_{j}}{2} \ket{\Phi_{j}^-} \right ).
\end{equation}
Similarly to Claim 2 in \cite{Ambainis:03a}, we have

\begin{Lemma}
If $\ket{w'_{\alpha}}$ is orthogonal to $\ket{\psi_{good}}$, then $|\omega_\alpha\ra =\ket{\psi_{good}}+
i\ket{w'_{\alpha}}$ is an eigenvector of $U'$ with eigenvalue
$e^{i\alpha}$ and $|\omega_{-\alpha}\ra =\ket{\psi_{good}}+ i\ket{w'_{-\alpha}}$ is an eigenvector
of $U'$ with eigenvalue $e^{-i\alpha}$.
\end{Lemma}

\begin{proof}
The proof is similar to \cite{Ambainis:03a}, but we include
it for completeness.

Apply $U'$ to $|\omega_\alpha\ra$ and expand in the eigenbasis of $U$:
\begin{equation}
\begin{array}{r}
U'|\omega_\alpha\ra = U(I-2|s,v\ra \la s,v|)
(\ket{\psi_{good}}+i\ket{w'_{\alpha}})=U(-\ket{\psi_{good}}+i\ket{w'_{\alpha}})= 
a_0 ( -1+i\cot \frac{\alpha}{2} ) \ket{\Phi_0} +\\
 \sum_{j} a_j \left( e^{i \theta_{j}}( -1+i\cot \frac{\alpha-\theta_{j}}{2} ) \ket{\Phi_{j}^+}
+  e^{-i \theta_{j}} ( -1+i\cot \frac{\alpha+\theta_{j}}{2} ) \ket{\Phi_{j}^-}  
\right).
\end{array} \nonumber
\end{equation}
In this equation, every coefficient is equal to the corresponding
coefficient in $e^{i\alpha} (\ket{\psi_{good}}+i\ket{w'_{\alpha}})$.
Namely, for the coefficient of $\ket{\Phi_0}$, we have
\[ \left( -1+i\cot\frac{\alpha}{2} \right) =
\frac{e^{i(\frac{\pi}{2}+\frac{\alpha}{2})}}{\sin\frac{\alpha}{2}} =
e^{i\alpha}
\frac{e^{i(\frac{\pi}{2}-\frac{\alpha}{2})}}{\sin\frac{\alpha}{2}}
= e^{i\alpha} \left( 1+i\cot\frac{\alpha}{2} \right) .\]
For the coefficient of $\ket{\Phi_{j}^+}$, we have
\[ e^{i\theta_{j}}\left( -1+i\cot\frac{-\theta_{j}+\alpha}{2} \right)=
e^{i\theta_{j}}
\frac{e^{i(\frac{\pi}{2}-\frac{\theta_{j}}{2}+\frac{\alpha}{2}) }}{\sin
\frac{-\theta_{j}+\alpha}{2}}=e^{i\alpha}
 \frac{e^{i(\frac{\pi}{2}+\frac{\theta_{j}}{2}-\frac{\alpha}{2}) }}{\sin
\frac{-\theta_{j}+\alpha}{2}} = e^{i\alpha}
\left( 1+i\cot\frac{-\theta_{j}+\alpha}{2} \right) \]
and, similarly, the 
conditions for the coefficients of $\ket{\Phi_{j}^-}$ are satisfied.
\end{proof}
By Eqs. (\ref{eq-psigood}) and (\ref{eq-a11}),
$\langle s,v|w'_{\alpha}\rangle=0$ is equivalent to
\begin{equation}
\label{eq-a1}
a_0^2 \cot \frac{\alpha}{2} + \sum_{j=1}^m a_j^2 (\cot
\frac{\alpha+\theta_{j}}{2}+\cot \frac{\alpha-\theta_{j}}{2}) =0 .
\end{equation}
Let $\theta_{min}$ be the smallest of $\theta_1$, $\ldots$, $\theta_m$.
Then, this equation has exactly one solution in
$[0, \theta_{min}]$ and one solution in $[-\theta_{min}, 0]$.
The reason for that is that the $\cot$ function is decreasing
(except for $x=k \pi$, where it goes to $-\infty$ for $x<k\pi$
and $+\infty$ for $x>k\pi$).
Therefore, the whole right hand side is decreasing,
except if one of $\frac{\alpha}{2}$,
$\frac{\alpha+\theta_{j}}{2}$, $\frac{\alpha-\theta_{j}}{2}$
becomes a multiple of $\pi$.
This happens for $\alpha=0$ and $\alpha=\pm \theta_{min}$.
Since $\theta_{min}$ is the smallest of $\theta_{j}$,
$[-\theta_{min}, 0]$ and $[0, \theta_{min}]$ contain
no values of $\alpha$ for which one of the $\cot$ becomes
infinity. On the interval $[0, \theta_{min}]$
the left-hand side of (\ref{eq-a1}) goes to $+\infty$
if $\alpha\rightarrow 0$, $-\infty$ if $\alpha\rightarrow\theta_{01}$
and is $0$ for exactly one value of $\alpha$ between $0$ and $\theta_{01}$.
This means that the two eigenvectors of $U'$ in the arc $\cal A$ are of
the form $|s,v\ra + i |\omega_\alpha '\ra$ with $|\omega_\alpha'\ra$ as
in Eq. (\ref{eq-a11}).

Next, let us determine this $\alpha$.
Since $$\cot x+\cot y=\frac{\cos x}{\sin x}+\frac{\cos y}{\sin y}=
\frac{\cos x \sin y+\cos y \sin x}{\sin x \sin y}=\frac{\sin(x+y)}{
\sin x \sin y}=2 \frac{\sin(x+y)}{\cos (x-y) -\cos (x+y)},$$ 
Eq. (\ref{eq-a1}) is equivalent to
$  a_0^2 \cot \frac{\alpha}{2} + \sum_{j} 2 a_j^2
\frac{\sin \alpha}{\cos \theta_{j} -\cos \alpha} =0$ which, in turn, is equivalent to
\[ a_0^2 \frac{\cot \frac{\alpha}{2}}{\sin\alpha}=
\sum_{j} 2 a_j^2 \frac{1}{\cos \alpha-\cos\theta_{j}} .\]
For $\alpha=o(1)$, $\sin\alpha=(1-o(1))\alpha$ and
$\cot\alpha=(1+o(1))\frac{1}{\alpha}$.
Therefore, we have, with $\cos\alpha\leq 1$
\begin{equation}
\label{eq-a2}
(1+o(1)) \frac{a^2_0}{\alpha^2} =
\sum_{j} a_j^2 \cdot \frac{1}{\cos \alpha-\cos\theta_{j}}  \geq
\sum_{j} a_j^2 \cdot \frac{1}{1-\cos\theta_{j}}.
\end{equation}

This implies 
\begin{equation} \label{eq:upperalpha}
 \alpha\leq \frac{1}{\sqrt{2}} 
\frac{1}{\sqrt{\sum_j \frac{a_j^2}{a_0^2} \frac{1}{1-\cos\theta_j}}} .
\end{equation}
It remains to lower bound $\alpha$.

Assume that $\alpha<\frac{1}{2}\theta_{min}$.
(Otherwise, the lower bound of the lemma is true.)
Then, we have 
\begin{equation}
\label{eq-half}
\cos \alpha-\cos \theta_{j} \geq \cos \frac{\theta_j}{2} - \cos \theta_j \geq
\frac{1}{2} (1-\cos \theta_{j}).
\end{equation}
The first inequality follows from $\alpha<\frac{\theta_{min}}{2}\leq \frac{\theta_j}{2}$
and $\cos$ being decreasing on $[0, \pi]$.
The second inequality is equivalent to $\cos \frac{\theta_j}{2} \geq \frac{1}{2} 
(1+\cos \theta_j)$ which follows from $1+\cos\theta_j=2\cos^2 \theta_j \leq 2 \cos \theta_j$.
Eq. (\ref{eq-a2}) and (\ref{eq-half}) together imply
\begin{equation} \label{eq-a2-2}
 (1+o(1)) \frac{a_0^2}{\alpha^2} \leq
\frac{1}{2} \sum_{j} a_j^2 \frac{1}{1-\cos\theta_{j}} ,
\end{equation}
which implies the lower bound on $\alpha$.
\end{proof}

\begin{proof} [ of Lemma \ref{lem:2}]
We will show that the starting state is close to 
the state $\ket{w_{start}}=\frac{1}{\|w'_{start}\|} \ket{w'_{start}}$, 
$\ket{w'_{start}}=\frac{1}{\sqrt{2}} \ket{w_{\alpha}}-
\frac{1}{\sqrt{2}} \ket{w_{-\alpha}}$.
By Eq. (\ref{eq-a11}), we have
\[ \ket{w'_{start}}=\sqrt{2} a_0 \cot \frac{\alpha}{2}  i \ket{\psi_{start}} +
\sum_j \sqrt{2} a_j \left( \cot \frac{\alpha+\theta_j}{2} - 
\cot \frac{\theta_j -\alpha}{2} \right) i 
\left( \ket{\Phi^+_j}-\ket{\Phi^-_j} \right) .\]
We have $\lbra \psi_{start} | w_{start}\rket 
= \frac{\lbra \psi_{start} | w'_{start}\rket}{\|w'_{start}\|} =
\frac{\sqrt{2} a_0 \cot \frac{\alpha}{2}}{\|w'_{start}\|}$.
Therefore, we need to bound $\| w'_{start} \|$.
We have
\[ \| w'_{start} \|^2 = 2 a_0^2 \cot^2 \frac{\alpha}{2} + 
4 \sum_j a_j^2 \left( \cot \frac{\alpha+\theta_j}{2} - \cot \frac{\theta_j -\alpha}{2} \right)^2 .\]
Since $\cot x=(1+o(1)) \frac{1}{x}$, 
the first term is $2(1+o(1)) a_0^2 \frac{4}{\alpha^2}=\Theta(a_0^2/\alpha^2)$.
Similarly to the previous lemma, we have
\[ \cot \frac{\alpha+\theta_j}{2} - \cot \frac{\theta_j -\alpha}{2} =
\frac{\sin \alpha}{\cos \alpha-\cos \theta_j} = \frac{(1+o(1)) \alpha}{\cos \alpha-\cos\theta_j} .\]
Since $\alpha<\frac{1}{2} \theta_{min}$, 
we have $\cos \alpha-\cos\theta_j \geq \frac{1}{2}(1-\cos \theta_j)$
(similarly to Eq. (\ref{eq-half})).
Therefore,
\[ \frac{\| w' \|^2}{a_0^2 \cot^2 \frac{\alpha}{2} } 
\leq 1+\frac{\sum_{j} a_j^2 \left( \frac{(1+o(1)) \alpha}{0.5 (1-\cos\theta_{j})^2 } \right)^2}{
\Theta(a_0^2/\alpha^2)} =
1+ \Theta\left(\alpha^2 \sum_j \frac{a_j^2}{a_0^2}  
\left( \frac{\alpha}{(-\cos\theta_{j}}\right)^2 \right).\]
This means that
\[ \lbra \psi_{start} | w\rket =
\frac{\sqrt{2} a_0 \cot \frac{\alpha}{2}}{\|w'\|} =
1- \Theta\left( \alpha^4 \sum_j \frac{a_j^2}{a_0^2}  
 \frac{1}{(1-\cos\theta_{j})^2} \right) .\]
\end{proof}

\begin{proof} [ of Lemma \ref{lem:3}]
Let $\ket{w_{good}}=\frac{1}{\sqrt{2}} \ket{w_{\alpha}} - 
\frac{1}{\sqrt{2}}\ket{w_{-\alpha}}$. 
We consider the unnormalized state 
$\ket{w'_{good}}=\ket{w'_{\alpha}} - \ket{w'_{-\alpha}}$.
Obviously, $\ket{w_{good}}=\frac{\ket{w'_{good}}}{\|w'_{good}\|}$.
We have 
\[ \ket{w'_{good}}=2 a_0\ket{\psi_{start}} + \]
\[ \sum_{j=1}^m a_j \left( ( 2+i \cot \frac{\alpha+\theta_j}{2}+i\cot \frac{-\alpha+\theta_j}{2} )
\ket{\psi_j^+}  + ( 2+i \cot \frac{\alpha-\theta_j}{2}+i\cot \frac{-\alpha-\theta_j}{2} )
\ket{\psi_j^-} \right) .\]
Also $\lbra \psi_{good} | w_{good}\rket = 
\frac{\lbra \psi_{good} | w'_{good}\rket}{\|w'_{good}\|}$.
Futhermore, $\lbra \psi_{good} | w'_{good} \rket = 2 a^2_0 + \sum_{j=1}^m 4 a^2_j 
= 2\|\ket{\psi_{good}}\|^2=2$.
(The imaginary terms cancel out because 
$\cot \frac{\pm\alpha+\theta_j}{2}=-\cot \frac{\mp\alpha-\theta_j}{2}$.)
It remains to bound $\|w'_{good}\|$.
We have
\[ \| w'_{good}\|^2 = 2 a^2_0 + \sum_{j=1}^m 4 a^2_j + 
\sum_{j=1}^m 2 a_j^2 \left( \cot \frac{\alpha+\theta_j}{2}
+i\cot \frac{-\alpha+\theta_j}{2} \right)^2 \]
\[ = 2 + \sum_{j=1}^m 2 a_j^2 \left(\cot \frac{\alpha+\theta_j}{2}+
\cot \frac{-\alpha+\theta_j}{2} \right)^2 
.\]
Since $\alpha<\frac{1}{2} \theta_{min}$, this sum is at most 
\[ 2+\sum_{j=1}^m 2 a_j^2 (2\cot \frac{\theta_j/2}{2})^2 \leq 
2+\Theta\left( \sum_{j=1}^m a_j^2 \cot^2 \frac{\theta_j}{4}\right) .\]
Therefore, $\|w'\|=\Theta(\max(\sqrt{\sum_j a_j^2 \cot^2 \frac{\theta_j}{4}},1))$
and  $\lbra \psi_{good} | w\rket =
\Theta\left(\min\left(\frac{1}{\sqrt{\sum_j a_j^2 \cot^2 \frac{\theta_j}{4}}}, 1\right) \right)$.
\end{proof}

\section{General graphs}

The approach and methods we have presented are amenable to analyze quantum walk algorithms 
on other graphs $G$. 
All we need is the eigenspectrum of the unperturbed walk $U$, an appropriate subspace $H'_0$ 
containing no $1$-eigenvectors of $U$ (which is equivalent to proving that all 
but one $1$-eigenvector of $U$ is orthogonal to $|s\ra$), 
and the sums in Lemmas \ref{lem:1}, \ref{lem:2} and \ref{lem:3} involving the eigenvalues of $U$ 
(which give the angle $\alpha$ and the overlaps). 
Then we can apply Lemmas \ref{lem:1}-\ref{lem:3} to get the desired result.
\paragraph{Hypercube:} For instance we can derive the performance of the random walk search 
algorithm on the hypercube, given in \cite{Shenvi:02b} without having to guess the form of 
the eigenvalues $|\pm \omega_0\ra$.  \cite{Shenvi:02b} showed that the random walk search 
algorithm after time $T=\frac{\pi}{2}\sqrt{N}$ gives a probability of $\approx \frac{1}{2}$ 
to measure the marked state. 
For the $d$-dimensional hypercube (with $N=2^d$ vertices), the transformation $U$ has 
$d2^d$ eigenvectors.
An argument similar to Claim \ref{claim:0} shows
that the quantum walk stays in the $2^d$-dimensional subspace spanned by $2^d$ eigenvectors
with eigenvalues 
$e^{i \theta_k}$ \cite{Moore:01a} with
$$\cos \theta_k = 1-\frac{2k}{d}$$
for $k=0 \ldots d$, each with degeneracy $\binom{d}{k}$. 
Among those, $\ket{\Phi_0}$ is the only eigenvector with eigenvalue 1, thus we have an instance of the abstract search.
We can now apply Lemmas \ref{lem:1}-\ref{lem:3}.
For Lemma \ref{lem:1}, we need to compute the sum of inverse gaps of all 
eigenvalues of $U$ in Lemma \ref{lem:thetasum}, which is now of the form 
$$ \sum_{k=1}^d \binom{d}{k} \frac{1}{1-\cos \theta_k}=\frac{d}{2}\sum_{k=1}^d \binom{d}{k} \frac{1}{k}=2^d(1+o(1))=N(1+o(1)).$$
This gives a time $T=\Theta(\sqrt{N})$ to rotate the state.
Evaluating the quantities in Lemmas \ref{lem:2} and \ref{lem:3}
shows that measuring the final state gives a marked location with constant
probability.

\paragraph{Remarks on general graphs}

We have seen that  crucial to ``good'' performance of these algorithms are essentially two ingredients:
\begin{enumerate}
\item{Coin property: The relevant Hilbert space $\H'_0$ of the perturbed walk $U'$ does not contain $1$ eigenvectors (i.e. all $1$-eigenvectors of $U$ except the starting state $|\Phi_0\ra$ are orthogonal to the marked state $|s,v\ra$).} 
\item{Graph property: The gap $g$ between the $1$-eigenvalue and the real part of the next closest eigenvalue of $U$ is sufficiently large (determines the overlap of the initial state with the two relevant eigenvectors). Furthermore the sum of inverse gaps of the eigenvalues of $U$, i.e. of terms $(1-\cos \theta)^{-1}$ where $\theta$ is an eigenvalue of $U$, is sufficiently small, such that the angle between the two eigenvectors of $U'$ with eigenvalue closest to one is sufficiently large (determines the speed of the algorithm).}
\end{enumerate}
We call the first item a ``Coin'' property because, as we have seen, it is the choice of coin that determines this behavior. We call the second property a ``Graph'' property because the gap and the closeness of the perturbed eigenvalues to $1$ depend on the topology of the graph. 

To be more precise let us carry our argument through for {\em Cayley-graphs} of Abelian groups.
The eigenvalues of the unperturbed walk $U$ are ``split'' by the coin to be ``around'' 
the eigenvalues of the normalized adjacency matrix of the graph. 
More precisely, note that the adjacency matrix of a Cayley graph can be written 
as a sum of commuting shift operations over all $d$ directions $A=\frac{1}{d}\sum_{i=1}^d S_i$. 
The eigenvalues of $A$ are just sums of eigenvalues of shifts 
(which are the Fourier coefficients). 
For instance in the case of the grid the eigenvalues of $A$ 
are of the form 
$\frac{1}{2}(\cos \frac{2 k \pi}{\sqrt{N}}  +\cos \frac{2 l \pi}{\sqrt{N}} )
=\frac{1}{4}(\omega^k+\omega^{-k}+\omega^l+\omega^{-l})$ for $k,l=0 \ldots \sqrt{N}-1$. 
From Eq. (\ref{eq:ckl}) we see that in general the coin ``changes'' the eigenvalues to be 
some linear combination of $\omega^{\pm k,l}$. 
This is what happens in general, and the resulting eigenspectrum will 
have gaps of the same order of magnitude as the spectrum of the matrix $A$ 
(which defines the simple random walk on the graph).
Thus, it is likely that the sum $\sum_{\theta} \frac{1}{1-\cos \theta}$ ($\theta$ eigenvalue of $U$) 
which determine the speed of the algorithm can be estimated from graph properties alone (given a successful choice of coin). 

\paragraph{Related work:}
Analyzing quantum walks on general graphs has been recently
considered by Szegedy \cite{Szegedy:04a} who has shown a following 
general result. If $\epsilon$ fraction of all vertices of a graph is
marked and all eigenvalues of a graph differ from
1 by at least $\delta$, then $O(\frac{1}{\sqrt{\epsilon\delta}})$ steps of
quantum walk suffice. This contributes to the same goal as
our paper: developing general tools for analyzing quantum walks
and using them in quantum algorithms.

The power of two methods (ours and \cite{Szegedy:04a}) 
seems to be incomparable. The strength of our method is that
it is able to exploit a finer structure of the graph.
For example, consider the 2-dimensional grid with a unique marked item. 
Applying the theorem of \cite{Szegedy:04a}
gives a running time of $O(N^{3/4})$,
compared to $O(\sqrt{N} \log N)$ for ours.
(One marked item is a $1/N$ fraction of all items and the eigenvalue
gap $\delta$ for the grid is $1/\sqrt{N}$.) 
The reason for this difference is that 
{\em most} eigenvalues are far from 1. 
The approach of \cite{Szegedy:04a} uses worst-case (minimum) difference
between 1 and eigenvalues of the graph, which is small ($\frac{1}{\sqrt{N}}$). 
Our approach uses a quantity ($\sum_{\theta} \frac{1}{1-\cos \theta}$)
that involves all eigenvalues, capturing the fact that most eigenvalues
are not close to 1.  

The strength of Szegedy's \cite{Szegedy:04a} analysis is that it allows 
to handle multiple marked locations with no extra effort, which
we have not been able to achieve with our approach. 
It might be interesting to combine the methods so that 
both advantages can be achieved at the same time.

\section{Conclusion}

We have shown that the discrete quantum walk can search the 2D grid in time $O(\sqrt{N}\log N)$ and
higher dimensional grids in time $O(\sqrt{N})$. 
This improves over previous search algorithm and shows an interesting
difference between discrete and continous time quantum walks.
More generally, we have opened the route to a general analysis of random walks 
on graphs by providing the necessary toolbox. 

The main open problems are applying this toolbox to other problems and
learning to analyze quantum walks if there are multiple (more than 2) 
marked locations. In the case of our problem, it is possible to reduce 
the multiple marked location case to the single location case at the cost of 
increasing the running time by a factor of $\log N$ \cite{Aaronson:03a}.
Still, it would be interesting to be able to analyze the multiple item case directly.
A recent paper by Szegedy \cite{Szegedy:04a} has shown how to
analyze the multiple solution case for a different quantum walk
algorithm, element distinctness \cite{Ambainis:03a}. It is open
whether the methods from \cite{Szegedy:04a} can be applied to search
on grids.

\paragraph{Acknowledgements}
Above all we thank Neil Shenvi for providing us with the results of his numerical simulations and stimulating us to search for an analytical proof.
JK thanks Fr\'ed\'eric Magniez and Oded Regev for helpful discussions.
AA is supported by NSF Grant DMS-0111298.
Part of this work was done while AA was at University of Latvia and 
University of California, Berkeley, supported by Latvia Science Council
Grant 01.0354 and DARPA and Air Force Laboratory, Air Force Materiel Command, USAF, under
agreement number F30602-01-2-0524, respectively.
JK acknowledges support by ACI S\'ecurit\'e Informatique, 2003-n24, projet
"R\'eseaux Quantiques" and by  DARPA
and Air Force Laboratory, Air Force Materiel Command, USAF, under
agreement number F30602-01-2-0524, and by  DARPA and the Office of Naval
Research under grant number FDN-00014-01-1-0826.
Any opinions, findings, conclusions or recommendations expressed in
this paper are those of authors and do not necessarily reflect
the views of funding agencies.


\begin{thebibliography}{BHMT02}  \bibitem[AA03]{Aaronson:03a} S.~Aaronson and A.~Ambainis. \newblock Quantum search of spatial regions. \newblock In {\em Proc. 44th Annual IEEE Symp. on Foundations of Computer   Science (FOCS)}, pages 200--209, 2003.  \bibitem[AAKV01]{Aharonov:01a} D.~Aharonov, A.~Ambainis, J.~Kempe, and U.~Vazirani. \newblock Quantum walks on graphs. \newblock In {\em Proc. 33th STOC}, pages 50--59, New York, NY, 2001. ACM.  \bibitem[ABN{\etalchar{+}}01]{Ambainis:01b} A.~Ambainis, E.~Bach, A.~Nayak, A.~Vishwanath, and J.~Watrous. \newblock One-dimensional quantum walks. \newblock In {\em Proc. 33th STOC}, pages 60--69, New York, NY, 2001. ACM.  \bibitem[Amb03]{Ambainis:03a} A.~Ambainis. \newblock Quantum walk algorithm for element distinctness, 2003. \newblock lanl-ar{X}ive quant-ph/0311001.  \bibitem[Ben02]{Benioff:02a} Paul Benioff. \newblock Space searches with a quantum robot. \newblock In {\em Quantum computation and information (Washington, DC, 2000)},   volume 305 of {\em Contemp. Math.}, pages 1--12. Amer. Math. Soc.,   Providence, RI, 2002.  \bibitem[BHMT02]{Brassard:02a} G.~Brassard, P.~H{\o}yer, M.~Mosca, and A.~Tapp. \newblock Quantum amplitude amplification and estimation. \newblock In {\em Quantum computation and information (Washington, DC, 2000)},   volume 305 of {\em Contemp. Math.}, pages 53--74. Amer. Math. Soc.,   Providence, RI, 2002.  \bibitem[CCD{\etalchar{+}}03]{Childs:02a} A.~M. Childs, R.~Cleve, E.~Deotto, E.~Farhi, S.~Gutmann, and D.~A. Spielman. \newblock Exponential algorithmic speedup by a quantum walk. \newblock In {\em Proc. 35th STOC}, pages 59--68, 2003. \newblock quant-ph/0209131.  \bibitem[CE03]{Childs:03b} A.M. Childs and J.M. Eisenberg. \newblock Quantum algorithms for subset finding. \newblock Technical report, lanl-ar{X}ive quant-ph/0311038, 2003.  \bibitem[CFG02]{Childs:01a} A.~Childs, E.~Farhi, and S.~Gutmann. \newblock An example of the difference between quantum and classical random   walks. \newblock {\em Quantum Information Processing}, 1:35, 2002. \newblock lanl-report quant-ph/0103020.  \bibitem[CG03]{Childs:03a} A.M. Childs and J.~Goldstone. \newblock Spatial search by quantum walk. \newblock Technical report, lanl-ar{X}ive quant-ph/0306054, 2003.  \bibitem[FG98]{Farhi:98a} E.~Farhi and S.~Gutmann. \newblock Quantum computation and decision trees. \newblock {\em Phys. Rev. A}, 58:915--928, 1998.  \bibitem[Gro96]{Grover:96a} L.~Grover. \newblock A fast quantum mechanical algorithm for database search. \newblock In {\em Proc. 28th STOC}, pages 212--219, Philadelphia, Pennsylvania,   1996. ACM Press.  \bibitem[Kem03a]{Kempe:03b} J.~Kempe. \newblock Quantum random walks - an introductory overview. \newblock {\em Contemporary Physics}, 44(4):302--327, 2003. \newblock lanl-ar{X}ive quant-ph/0303081.  \bibitem[Kem03b]{Kempe:02b} J.~Kempe. \newblock Quantum walks hit exponentially faster. \newblock In {\em RANDOM-APPROX 2003}, Lecture Notes in Computer Science, pages   354--369, Heidelberg, 2003. Springer. \newblock lanl-ar{X}iv quant-ph/0205083.  \bibitem[Mey96]{Meyer:96a} D.~Meyer. \newblock From quantum cellular automata to quantum lattice gases. \newblock {\em J. Stat. Phys.}, 85:551--574, 1996.  \bibitem[MR95]{Motwani:book} R.~Motwani and P.~Raghavan. \newblock {\em \em Randomized Algorithms}. \newblock Cambridge University Press, 1995.  \bibitem[MR02]{Moore:01a} C.~Moore and A.~Russell. \newblock Quantum walks on the hypercube. \newblock In J.D.P. Rolim and S.~Vadhan, editors, {\em Proc. RANDOM 2002},   pages 164--178, Cambridge, MA, 2002. Springer.  \bibitem[MSS03]{Magniez:03a} F.~Magniez, M.~Santha, and M.~Szegedy. \newblock Quantum algorithms for the triangle problem. \newblock Technical report, 2003. \newblock lanl-ar{X}ive quant-ph/0310134. 
\bibitem[SKW03]{Shenvi:02b} N.~Shenvi, J.~Kempe, and K.B. Whaley. \newblock A quantum random walk search algorithm. \newblock {\em Phys. Rev. A}, 67(5):052307, 2003. \newblock lanl-ar{X}ive quant-ph/0210064. 
\bibitem[She03]{Shenvi:unp} Neil Shenvi. \newblock Random Walk Simulations. \newblock unpublished.
\bibitem[Sze04]{Szegedy:04a}
M. Szegedy. 
Spectra of quantized walks and a $\sqrt{\delta\epsilon}$ rule,
quant-ph/0401053.
 \end{thebibliography}

\newcommand{\etalchar}[1]{$^{#1}$}

\end{document}